\documentclass[preprint]{aastex}
\usepackage{amsmath}%
\usepackage{amsfonts}%
\usepackage{amssymb}%
\usepackage{graphicx}
\usepackage[english]{babel}
\usepackage[latin1]{inputenc}
\usepackage{multicol}
\usepackage{natbib}

\begin{document}
\bibliographystyle{abbrvnat}
\title{Parasitic interference in long baseline optical interferometry - Requirements for hot Jupiter-like planet detection}
\author{A. Matter\altaffilmark{1}, B. Lopez, S. Lagarde}
\affil{Laboratoire Fizeau, UMR 6525, UNS - Observatoire de la côte d'azur, BP 4229, F-06304 Nice Cedex 4, France} 
\author{W.C. Danchi}
\affil{NASA/GSFC, Greenbelt, MD 20771, USA}
\author{S. Robbe-Dubois, R.G. Petrov} 
\affil{Laboratoire Fizeau, UMR 6525, UNS - Observatoire de la côte d'azur, 06108 Nice cedex 02, France}
\and
\author{R. Navarro}
\affil{NOVA-ASTRON, P.O. Box 2, 7990 AA Dwingeloo, The Netherlands}
\altaffiltext{1}{Corresponding author: matter@oca.eu} 


\begin{abstract}
The observable quantities in optical interferometry, which are the modulus and the phase of the complex visibility, may be corrupted by parasitic fringes superimposed on the genuine fringe pattern. These fringes are due to an interference phenomenon occurring from straylight effects inside an interferometric instrument. We developed an analytical approach to better understand this phenomenon when straylight causes crosstalk between beams.\\
We deduced that the parasitic interference significantly affects the interferometric phase and thus the associated observables including the differential phase and the closure phase. The amount of parasitic flux coupled to the piston between beams appears to be very influential in this degradation. For instance, considering a point-like source and a piston ranging from $\lambda/500$ to $\lambda/5$ in L band ($\lambda=3.5\:\mu$m), a parasitic flux of about 1\% of the total flux produces a parasitic phase reaching at most one third of the intrinsic phase. The piston, which can have different origins (instrumental stability, atmospheric perturbations, ...), thus amplifies the effect of parasitic interference.\\
According to specifications of piston correction in space or at ground level (respectively $\lambda/500\approx 2$nm and $\lambda/30\approx 100$nm), the detection of hot Jupiter-like planets, one of the most challenging aims for current ground-based interferometers, limits parasitic radiation to about 5\% of the incident intensity. This was evaluated by considering different types of hot Jupiter synthetic spectra. 
Otherwise, if no fringe tracking is used, the detection of a typical hot Jupiter-like system with a solar-like star would admit a maximum level of parasitic intensity of 0.01\% for piston errors equal to $\lambda$/15. If the fringe tracking specifications are not precisely observed, it thus appears that the allowed level of parasitic intensity dramatically decreases and may prevent the detection. In parallel, the calibration of the parasitic phase by a reference star, at this accuracy level, seems very difficult. Moreover, since parasitic phase is an object-dependent quantity, the use of a hypothetical phase abacus, directly giving the parasitic phase from a given parasitic flux level, is also impossible. Some instrumental solutions, implemented at the instrument design stage for limiting or preventing this parasitic interference, appears to be crucial and are presented in this paper.
\end{abstract}


\section{Introduction}
Stellar optical interferometry has substantially evolved in terms of instrument and operation since Fizeau's idea to use this technique for measuring the diameter of stars \citep{Fizeau1868}.
Since Michelson's interferometer, where separated mirrors were fixed on the same telescope mounting \citep{1920ApJ....51..257M}, interferometry with independent telescopes has allowed access to very long baselines \citep{1975ApJ...196L..71L} and consequently high angular resolution. 
In this case, which corresponds to most current interferometers, the beams coming from each telescope, are carried through tunnels up to a combining device. The beams are often reduced in size for practical reasons, and the current instruments are characterized by optical modules performing various functions such as spatial filtering, spectral band separation, and spectral resolution. The transport of these beams through multiple optical modules can be problematic.\\ 
A parasitic interference may occur because of diffraction effects associated with beam propagation along finite size optics. This diffraction can then produce a cross-talk between beams. 
The result is the superposition on the detector of several fringes systems having different phases. Therefore the interferometric signal of the independent beams is affected. While the intrinsic interferometric fringe pattern corresponds to Fizeau-like fringes, the parasitic interference induces two further components to the signal: Fizeau-like mirror fringes and Young-like fringes. The latter contribution is independent of the object position in the sky. We will develop this description in the article.\\
This interference corrupts the value of interferometric observables : modulus and phase of the complex visibility. The understanding of this phenomenon is of general interest for optical long baseline interferometers which combine multiple beams and often require some "compactness" in the opto-mechanical elements.\\
In Section 2, a simple formalism is developed, explicitly showing the different contributions of the parasitic interference. In Section 3, the resulting interferometric observables are written for the general case of an extended source. Then the theoretical cases of an unresolved source and of a stellar system with a hot Jupiter are highlighted. Sections 4 and 5 quantify the impact of the parasitic interference on the theoretical phase signal of these types of sources. The parameters involved in this quantitative study are the factor of parasitic flux between beams, the piston, and the photometric imbalance between interferometric arms. The impact of the parasitic interference on hot Jupiter detection is evaluated by using different synthetic spectra of such extrasolar planets. Requirements on the straylight level limits are also given.

\section{Formalism of the problem}
\subsection{Overview}
In general, long-baseline stellar interferometry consists of sampling an incident wave packet by means of telescopes at different locations. These coherent waves are combined and the resulting interference pattern is extracted in order to measure the complex degree of coherence of the radiation field. Then the brightness distribution of the source can be recovered thanks to the Van Cittert and Zernike theorem. This complex degree of coherence corresponds to the covariance of the electric fields collected by the telescopes.
If a parasitic interference occurs between the collection and the recombination steps, the 'intrinsic' coherence between beams and consequently the resulting interferometric observables will be perturbed.\\
To our knowledge, the issue of parasitic interference has never been formalized except for the digital wave-front measuring interferometry technique. This well-established technique, described in \citet{1974ApOpt..13.2693B}, allows testing of optical surfaces and lenses, and measurement of wave-front deviations in the $\lambda/100$ range. Some systematic error sources such as 'extraneous' fringes, that is parasitic interference, were examined in the same paper and more extensively in \citet{Schwider:83}.
In our case, the theoretical description of this perturbation is hereafter detailed in the framework of a simple two-telescope interferometer.     

\subsection{Interferometric framework} 

To create a model of parasitic interference, we use a two-telescope interferometer characterized by a multi-axial scheme and an image plane recombination. Fig. 1 gives an illustration of such an interferometer observing an unresolved astrophysical source. For a general description of the output response of a single baseline interferometer, see also \citet{2007ApJ...657.1178E}.\\
First we write the complex amplitudes collected by both telescopes, i.e. the two samples of the wavefront, that we respectively note $\psi_1$ and $\psi_2$. We multiply each of them by a real transmission factor noted t$_1$ and t$_2$, which represents the transmission of electrical fields through the instrument.\\ 
Let us define $\epsilon$ to be the main fraction of $\psi_1$ propagating towards the path 1, and $\epsilon'$ to be the small fraction propagating towards the path 2. We assume the same parasitic effect for $\psi_2$. Fig. 2 shows two examples of cross-talk between beams. A possible imbalance of parasitic flux between paths, not presented in this paper, was considered in the study but appeared to be a second order parameter.
All the cross-talk occuring inside the instrument produces a resulting parasitized pattern in a conjugate of the pupil plane, located just before the recombination on the detector. This is described in the following equation giving the complex amplitude preceding the recombination :
 \begin{equation}
 \psi_{\rm pup}(x,y)=[\epsilon t_1\psi_1+\epsilon' t_2\psi_2]{\rm P}(x-\frac{b}{2},y) + [\epsilon t_2\psi_2+\epsilon' t_1\psi_1]{\rm P}(x+\frac{b}{2},y).
 \end{equation}
 Here $x$ and $y$ are the coordinates in the pupil plane. $b$ is the distance between the pupils of both telescopes, which are reduced in size at the entrance of the interferometer, $D$ being the diameter of these pupils. Since in a Michelson configuration the collection and correlation steps are separated, $b$, the 'reduced' baseline which forms the fringe pattern, is different from $B$, the real interferometric baseline which samples the incident wavefront. ${\rm P}(x,y)=\Pi(\frac{\sqrt{x^2+y^2}}{D})$, where $\Pi(x,y)$ is the transmission function of a circular aperture with a uniform transmission of 1 inside and 0 outside. From Eq.(1), we discuss how parasitic fringes are formed on the detector.\\
As we will see later, the complex degree of coherence between both parasitized beams will become $<[\epsilon t_1\psi_1+\epsilon' t_2\psi_2][\epsilon t_2\psi_2+\epsilon' t_1\psi_1]^*>$, instead of the 'intrinsic' coherence term $<\psi_1\psi_2^*>$. 

 \subsection{Interference and formation of parasitic fringes}  
In order to describe the formation of parasitic fringes, we consider the most general case of the observation of an extended source.
 The vector $\boldsymbol{\alpha}$ is the angular coordinate in the plane of the sky. The telescopes are located at $\boldsymbol{\gamma_1}$ and $\boldsymbol{\gamma_2}$, $\boldsymbol{\gamma}$ being the coordinate in the plane containing the telescopes and counted in units of wavelength ($\boldsymbol{\gamma}=\frac{\boldsymbol{{\rm r}}}{\lambda}$) (see Fig. 1). $\frac{\boldsymbol{B}}{\lambda}=\boldsymbol{\gamma_1}-\boldsymbol{\gamma_2}$ with $\boldsymbol{B}$ the interferometer baseline.
The electric field emitted by each point of the extended source, located at $\boldsymbol{\alpha}$, is represented by its amplitude A($\alpha$) and its phase $\phi(\boldsymbol{\alpha},{\rm t})$ :
\begin{mathletters}
\begin{equation}
\psi_{\rm S}(\boldsymbol{\alpha})={\rm A}(\boldsymbol{\alpha})e^{i[\omega t+\phi(\boldsymbol{\alpha},t)]}{\rm d}\boldsymbol{\alpha}.
\end{equation}
In the plane of the telescopes, the phase shift of the wavefront emitted by each point of the source and measured on the $\boldsymbol{\gamma}$ position is $\Delta\Phi=-2\pi\boldsymbol{\alpha}\cdot\boldsymbol{\gamma}$.
Therefore, when considering the contributions of all the emitting points of the source, each telescope observes a packet of wavefronts (respectively $\psi_1$ and $\psi_2$) : 
\begin{eqnarray}
\psi_1 & =\int {\rm A}(\boldsymbol{\alpha})e^{i[\omega t+\phi(\boldsymbol{\alpha},t)]}e^{i2\pi\boldsymbol{\alpha}\cdot\boldsymbol{\gamma_1}}d\boldsymbol{\alpha},  \\
\psi_2 & =\int {\rm A}(\boldsymbol{\alpha})e^{i[\omega t+\phi(\boldsymbol{\alpha},t)]}e^{i2\pi\boldsymbol{\alpha}\cdot\boldsymbol{\gamma_2}}d\boldsymbol{\alpha}.
\end{eqnarray}
W can now write the expression of the complex amplitude in the detector plane by performing the Fourier transform of Eq.(1) with respect to $x$ and $y$. In the detector plane, where $\beta$ and $\eta$ are the conjugate angular coordinates linked to $x$ and $y$, and $\hat{\rm P}(\beta,\eta)$ is the pupil diffraction function, we obtain :
\end{mathletters}
 {\small
  \begin{align}
 W(\beta,\eta) & =(\epsilon t_1\psi_1+\epsilon' t_2\psi_2)\hat{\rm P}(\beta,\eta)e^{\frac{-i\pi b\beta}{\lambda}}+(\epsilon t_2\psi_2+\epsilon' t_1\psi_1)\hat{\rm P}(\beta,\eta)e^{\frac{i\pi b\beta}{\lambda}}=X_1+X_2.
 \end{align}}$(\epsilon t_1\psi_1+\epsilon' t_2\psi_2)$ and $(\epsilon t_2\psi_2+\epsilon' t_1\psi_1)$ are not affected by the Fourier transform since $\psi_1$ and $\psi_2$ can be considered, in a good approximation, as constant over the aperture telescope area. This assumption is equivalent to the requirement that the source is not resolved by the telescopes themselves or that the field of view is small. In Eq.(3), $X_1=(\epsilon t_1\psi_1+\epsilon' t_2\psi_2)\hat{\rm P}(\beta,\eta)e^{\frac{-i\pi b\beta}{\lambda}}$ and $X_2=(\epsilon t_2\psi_2+\epsilon' t_1\psi_1)\hat{\rm P}(\beta,\eta)e^{\frac{i\pi b\beta}{\lambda}}$ represent the Fourier transform of the parasitized complex amplitudes of each beam. On the detector, we observe the following intensity pattern : 
\begin{mathletters} 
\begin{equation}
I(\beta,\eta)=<|X_1|^2>+<|X_2|^2>+2{\rm Re}<X_1X^*_2>.
\end{equation}
The photometric terms, $<|X_1|^2>$ and $<|X_2|^2>$, are equal to:
\begin{eqnarray}
\nonumber
<|X_1|^2>&=&\hat{\rm P}^2(\beta,\eta)<|\epsilon t_1\psi_1+\epsilon' t_2\psi_2|^2> \\
\nonumber
         &=&\hat{\rm P}^2(\beta,\eta)[\epsilon^2t_1^2<|\psi_1|^2>+\epsilon'^2t_2^2<|\psi_2|^2>+2\epsilon\epsilon't_1t_2Re<\psi_1\psi^*_2>] \\
				 &=&\hat{\rm P}^2(\beta,\eta)[(\epsilon^2t_1^2+\epsilon'^2t_2^2)\hat{\rm O}(0)]+2\epsilon\epsilon't_1t_2Re(\hat{\rm O}(\boldsymbol{\gamma_1}-\boldsymbol{\gamma_2})),
\end{eqnarray}
and
\begin{eqnarray}
<|X_2|^2>&=&\hat{\rm P}^2(\beta,\eta)[(\epsilon^2t_2^2+\epsilon'^2t_1^2)\hat{\rm O}(0)]+2\epsilon\epsilon't_1t_2Re(\hat{\rm O}(\boldsymbol{\gamma_2}-\boldsymbol{\gamma_1}))].
\end{eqnarray}Here ${\rm O}(\boldsymbol{\alpha})={\rm A}^2(\boldsymbol{\alpha})$ is the intensity distribution of the source. The Fourier transform of the intensity distribution of the source at the spatial frequency ($\boldsymbol{\gamma_1}-\boldsymbol{\gamma_2}$), also called the complex degree of mutual coherence, is : $\hat{{\rm O}}(\boldsymbol{\gamma_1}-\boldsymbol{\gamma_2})=<\psi_1\psi^*_2>=\int{{\rm O}(\boldsymbol{\alpha})e^{-i2\pi\boldsymbol{\alpha}\cdot(\boldsymbol{\gamma_1}-\boldsymbol{\gamma_2})}{\rm d}\boldsymbol{\alpha}}$. In this covariance operation, one of the integrals disappears because astrophysical sources are spatially incoherent when time averaged. We also note that $\hat{\rm O}(0)=<|\psi_1|^2>=<|\psi_2|^2>=\int{\rm O(\boldsymbol{\alpha})d\boldsymbol{\alpha}}$ is the total energy radiated by the source and collected by each telescope.\\
The term O$(\boldsymbol{\alpha})$ is a real function, so its Fourier transform, noted $\hat{{\rm O}}(\boldsymbol{\gamma_1}-\boldsymbol{\gamma_2})=\rho_{12}{\rm e}^{i\Phi_{12}}$, is a complex function with an even real part and an odd imaginary part. 
Finally, the addition of both photometric terms gives :
 \begin{equation}
 <|X_1|^2>+<|X_2|^2>=\hat{\rm P}^2(\beta,\eta)[(t_1^2+t_2^2)(\epsilon^2+\epsilon'^2)\hat{\rm O}(0)+4\epsilon\epsilon't_1t_2\rho_{12}\cos(\Phi_{12})].
 \end{equation}
In order to highlight the different fringe patterns appearing in the interferogram, let us explicitly show the real part of the correlation term, $<{\rm X}_1{\rm X}^*_2>$, containing the coherent flux and the spatial modulation :
  {\footnotesize \begin{eqnarray}
  \nonumber 
  {\rm Re}<{\rm X}_1{\rm X}^*_2>&=&{\rm Re}(<\hat{\rm P}^2(\beta,\eta)[(\epsilon t_1\psi_1+\epsilon' t_2\psi_2)e^{\frac{-i\pi b\beta}{\lambda}}][(\epsilon t_2\psi_2^*+\epsilon' t_1\psi_1^*)e^{\frac{-i\pi b\beta}{\lambda}}]>) \\
                                &=&\hat{\rm P}^2(\beta,\eta){\rm Re}[[\epsilon\epsilon'(t_1^2+t_2^2)\hat{\rm O}(0)+\epsilon^2t_1t_2\hat{{\rm O}}(\boldsymbol{\gamma_1}-\boldsymbol{\gamma_2})+\epsilon'^2t_1t_2\hat{{\rm O}}^*(\boldsymbol{\gamma_1}-\boldsymbol{\gamma_2})]e^{-2i\pi b\frac{\beta}{\lambda}}] \\
  \nonumber                              
                                &=&\hat{\rm P}^2(\beta,\eta)[\epsilon^2t_1t_2\rho_{12}\cos(\frac{2\pi b\beta}{\lambda}+\Phi_{12})+\epsilon\epsilon'(t_1^2+t_2^2)\hat{\rm O}(0)\cos(\frac{2\pi b\beta}{\lambda})+\epsilon'^2t_1t_2\rho_{12}\cos(\frac{2\pi b\beta}{\lambda}-\Phi_{12})]. 
  \end{eqnarray}}In this expression, each cosine modulation factor has a physical meaning. The first modulation in $\cos(\frac{2\pi}{\lambda}b\beta+\Phi_{12})$ is the intrinsic fringe pattern. The second modulation in $\cos(\frac{2\pi}{\lambda}b\beta)$ corresponds to Young fringes created by the interference between each of the beams and their corresponding diffracted part. The position of these fringes is fixed and does not depend on the object position in the sky. The last modulation in $\cos(\frac{2\pi}{\lambda}b\beta-\Phi_{12})$ is due to the interference between the diffracted part of both beams. Mirror-like fringes are thus created with an opposite phase with regard to the intrinsic fringe pattern. Fig. 1 shows these different fringe patterns in the case of an unresolved source. Each pattern has a different amplitude depending on the nature of the object and the amount of parasitic flux.
\end{mathletters}
\section{Highlight of the interferometric observables}
 \subsection{Resolved source}
 After describing the different fringe patterns due to the parasitic interference, we highlight the resulting interferometric observables that have been degraded. From the general correlation term $<{\rm X}_1{\rm X}^*_2>$, we extract the resulting parasitized coherent flux $\rho_{\rm ext}$ and parasitized phase $\chi_{\rm ext}$ :
 \begin{mathletters}
 \begin{align}
 \nonumber
 <{\rm X}_1{\rm X}^*_2> 
 \nonumber
 &=\hat{\rm P}^2(\beta,\eta)[\epsilon\epsilon'(t_1^2+t_2^2)\hat{\rm O}(0)+\epsilon^2t_1t_2\hat{{\rm O}}(\boldsymbol{\gamma_1}-\boldsymbol{\gamma_2})+\epsilon'^2t_1t_2\hat{{\rm O}}(\boldsymbol{\gamma_2}-\boldsymbol{\gamma_1})]e^{-2i\pi b\frac{\beta}{\lambda}} \\
 &=\hat{\rm P}^2(\beta,\eta)[\rho_{\rm ext}\:e^{i\chi_{\rm ext}}]\:e^{-2i\pi b\frac{\beta}{\lambda}}=\hat{\rm P}^2(\beta,\eta)\rho_{\rm ext}\:e^{-i(2\pi b\frac{\beta}{\lambda}-\chi_{\rm ext})}.
 \end{align}
 The resulting interferogram in the detector plane is :
 {\small
 \begin{align}
 \nonumber
 I(\beta,\eta) 
 &= \hat{\rm P}^2(\beta,\eta)[(t_1^2+t_2^2)(\epsilon^2+\epsilon'^2)\hat{\rm O}(0)+4\epsilon\epsilon't_1t_2\rho_{12}\cos(\Phi_{12})]+2\hat{\rm P}^2(\beta,\eta)\rho_{\rm ext}\cos(2\pi b\frac{\beta}{\lambda}-\chi_{\rm ext}) \\
 &= I_{0,{\rm ext}}[1+V_{\rm ext}\cos(2\pi b\frac{\beta}{\lambda}-\chi_{\rm ext})],
 \end{align}}
 where
 {\footnotesize 
 \begin{align}
 I_{0,{\rm ext}} &=\hat{\rm P}^2(\beta,\eta)[(t_1^2+t_2^2)(\epsilon^2+\epsilon'^2)\hat{\rm O}(0)+4\epsilon\epsilon't_1t_2\rho_{12}\cos(\Phi_{12})], \\
 V_{\rm ext} &=\frac{2\hat{\rm P}^2(\beta,\eta)\rho_{\rm ext}}{I_{0,{\rm ext}}}=\frac{2\sqrt{[\epsilon''(1+t_{12}^2)\hat{\rm O}(0)+t_{12}(1+\epsilon''^2)\rho_{12}\cos(\Phi_{12})]^2+[t_{12}(1-\epsilon''^2)\rho_{12}\sin(\Phi_{12})]^2}}{(1+\epsilon''^2)(1+t_{12}^2)\hat{\rm O}(0)+4\epsilon''t_{12}\rho_{12}\cos(\Phi_{12})}, \\
 \chi_{\rm ext} &= \arctan[\frac{t_{12}(1-\epsilon''^2)\sin(\Phi_{12})}{\epsilon''(1+t_{12}^2)\hat{\rm O}(0)+t_{12}(1+\epsilon''^2)\rho_{12}\cos(\Phi_{12})}].
 \end{align}}t$_{12}=\frac{{\rm t}_1}{{\rm t}_2}$ is the transmission ratio between both arms. $\epsilon"=\frac{\epsilon'}{\epsilon}$ represents the percentage of parasitic contribution, evaluated with respect to $\epsilon$ which is the fraction of the electric field passing through the right path. $\epsilon"^2=(\frac{\epsilon'}{\epsilon})^2$ is the equivalent ratio in terms of flux (or intensity), later called 'parasitic flux factor'. $I_{\rm 0,ext}$ is the photometric factor containing the parasitic contribution. Similarly, V$_{\rm ext}$ and $\chi_{\rm ext}$ are the parasitized visibility and phase. As a result, the parasitic phase, representing the parasitic contribution added to the intrinsic phase $\Phi_{12}$, is $\chi_{\rm ext}-\Phi_{12}$.\\
To summarize, the photometry, visibility and phase of a resolved source are degraded. This degradation depends on the complex degree of coherence of the source expressed by $\rho_{12}$ and $\Phi_{12}$. The coherent flux $\rho_{\rm ext}$ can not be calibrated by using the parasitized photometry even without any photometric imbalance (t$_{12}=1$). It also appears that the phase can not be calibrated. 
\end{mathletters}
\subsection{Unresolved source}
In this subsection let us consider the parasitic interference effect while observing a point-like source. The unresolved object is described by its angular position in the sky noted $\boldsymbol{\alpha_{\rm point}}$, with respect to the line of sight, and its monochromatic flux noted I$_{\rm point}(\lambda)$. The chromatic brightness distribution can be represented in the same way as in the resolved source case, such that only one tilted wavefront originates from the source :
\begin{mathletters}
\begin{equation}
{\rm O}_{\rm point}(\boldsymbol{\alpha},\lambda)=I_{\rm point}(\lambda)\delta(\boldsymbol{\alpha}-\boldsymbol{\alpha_{\rm point}}).
\end{equation}
$\hat{O}_{\rm point}({\rm \boldsymbol{\omega}},\lambda)$ is the Fourier transform of the brightness distribution, in the plane of spatial frequency $\boldsymbol{\omega}({\rm u},{\rm v})$ covered by a single-baseline interferometer. The u-axis is chosen along the interferometric baseline defined above and takes the value $\boldsymbol{{\rm u}}=\frac{\boldsymbol{{\rm B}}}{\lambda}$. Therefore we have :
\begin{equation}
\hat{O}_{\rm point}({\rm \boldsymbol{u}},\lambda)={\rm I}_{\rm point}(\lambda)e^{i2\pi \boldsymbol{{\rm u}\cdot \alpha_{\rm point}}}.
\end{equation}  
The fringe phase is the argument of $\hat{O}_{\rm point}({\rm \boldsymbol{u}},\lambda)$ : 
\begin{equation}
\Phi_{\rm point}(\lambda)=2\pi \boldsymbol{{\rm u}\cdot \alpha_{\rm point}}.
\end{equation}
Normally, $\Phi_{\rm point}(\lambda)$ should be defined with an additionnal unknown constant. The reason is that the measurement of the interferometric phase is affected by an ambiguity on the zero-delay point, that is the origin of the fringes, and by an unknown integer number of $2\pi$ phase rotation. Therefore the phase can only be addressed in a relative point of view, either between the source and a reference object, or between several spectral channels. The latter constitutes the differential approach which allows to remove the phase ambiguity thanks to a colour difference with an appropriate reference wavelength. $\Phi_{\rm point}(\lambda)$ is considered to be such an unambiguous differential phase, where the phase reference has been taken as the phase averaged on all the spectral channels of the observing band.\\     
Next the coherent flux is the amplitude of $\hat{O}_{\rm point}({\rm \boldsymbol{u}},\lambda)$ :
\begin{equation}
\rho_{\rm point}(\lambda)={\rm I}_{\rm point}(\lambda).
\end{equation}
The parasitized photometry, visibility, and phase of an unresolved source are deduced by replacing $\Phi_{12}$, $\rho_{12}$ and $\hat{O}(0)$ in Eqs.()()()
Replacing $\Phi_{12}$, $\rho_{12}$ and $\hat{O}(0)$ by $\Phi_{\rm point}(\lambda)$, $\rho_{\rm point}(\lambda)$ and ${\rm I}_{\rm point}(\lambda)$ in Eq.(5c), Eq.(5d) and Eq.(5e), allows us to deduce the parasitized photometry, visibility and phase of an unresolved source\footnote{The $\lambda$ dependency of $\Phi_{\rm point}(\lambda)$, $\rho_{\rm point}(\lambda)$ and ${\rm I}_{\rm point}(\lambda)$ has been removed in Eq.(6e), Eq.(6f), and Eq.(6g), for lightening the notations.} : 
{\footnotesize
\begin{align}
I_{\rm 0,point}(\lambda) & =I_{\rm point}\hat{\rm P}^2(\beta,\eta)[(\epsilon^2+\epsilon'^2)(t_1^2+t_2^2)+4\epsilon\epsilon't_1t_2\cos(\Phi_{\rm point})], \\
V_{\rm point}(\lambda) & =2\frac{\sqrt{(1+\epsilon''^2)^2t_{12}^2-4\epsilon''^2t_{12}^2\sin^2(\Phi_{\rm point})+\epsilon''^2(1+t_{12}^2)^2+2t_{12}(1+\epsilon''^2)\epsilon''(1+t_{12}^2)\cos(\Phi_{\rm point})}}{(t_{12}^2+1)(\epsilon''^2+1)+4\epsilon''t_{12}\cos(\Phi_{\rm point})},\\ 
\chi_{\rm point}(\lambda) & =\arctan(\frac{\sin(\Phi_{\rm point})(1-\epsilon''^2)t_{12}}{t_{12}(1+\epsilon''^2)\cos(\Phi_{\rm point})+\epsilon''(1+t_{12}^2)}).
\end{align}}Here $\boldsymbol{\gamma_1}-\boldsymbol{\gamma_2}=\frac{\boldsymbol{B}}{\lambda}$ and $\boldsymbol{\alpha}$ are considered to be colinear so we have : $\Phi_{\rm point}(\lambda)=\frac{2\pi B\alpha}{\lambda}$.
Regardless of the angular position of the star in the sky, other sources can contribute to the phase term, $\Phi_{\rm point}(\lambda)$, such as instrumental and/or atmospheric perturbations. As shown in Fig. 1, the total piston noted L and resulting from these different causes, is first corrected by the delay lines with an OPD model depending only on the angular distance between the source and the zenith. Then the remaining piston, noted by $\Delta$ in Fig. 1, may be corrected by a fringe tracker which commands the delay lines to adjust to a value of $\Delta$ with an accuracy defined by the fringe tracking specifications. This accuracy is noted $\delta$ and represents the residual piston that will affect the fringes within the signal envelope. Therefore the generalized expression of the phase term will be : $\Phi_{point}(\lambda)=\frac{2\pi \delta}{\lambda}$. Similarly to the resolved case, we consider $\chi_{\rm point}(\lambda)-\Phi_{\rm point}(\lambda)$ as the parasitic phase. 
\end{mathletters} \\
According to Eq.(6f), $V_{\rm point}=1$ if t$_{12}$=1, and depends neither on $\epsilon''$ and $\epsilon''^2$, nor on the piston. In the case of an unresolved source, the degradation due to the parasitic light can thus be corrected if we divide the parasitized coherent flux by the parasitized photometry. Though, this correction is effective only if the photometric imbalance t$_{12}$ is close or equal to 1.   
\subsection{Hot Jupiter-like source}
The direct detection of Earth-like extrasolar planets and the determination of their atmospheric features by means of stellar interferometry is a very challenging objective. Nowadays the observation of hot Jupiter-like planets constitutes an intermediate step in terms of difficulty. These planets are giant gaseous planets similar in size to Jupiter except that they orbit very close to their star. This proximity provides planet temperatures warm enough to allow significant infrared excess. The planetary flux can be decomposed in this way :
\begin{equation}
f_{\rm pl}(\lambda)=f_{\rm pl,refl}(\lambda)+f_{\rm pl,abs}(\lambda)+f_{\rm pl,int}(\lambda),
\end{equation}
where $f_{\rm pl,refl}(\lambda)$ is the component of the stellar light reflected by the planet, f$_{\rm pl,abs}(\lambda)$ the absorbed and thermally re-emitted component, and f$_{\rm pl,int}(\lambda)$ the intrinsic contribution of the planet. For non-irradiated objects, the main contribution would be limited to the intrinsic flux $f_{\rm pl,int}=\sigma T_{\rm int}^4$, where $f_{\rm pl,int}$ is now the bolometric intrinsic flux and $T_{\rm int}$ is the intrinsic temperature of the planet ($T_{\rm int}=T_{\rm eff}$ in this case).
For an irradiated planet, T$_{\rm eff}$ is not relevant because it is difficult to separate photons which are thermally radiated by the planet from those of the star itself reflected by the planet. Another temperature, describing the equlibrium state of the planet's dayside and separated from the reflected component, is thus defined : $\sigma T_{\rm eq}^4=\sigma T_{\rm int}^4+f_{\rm pl,abs}$, where $f_{\rm pl,abs}$ is a bolometric flux.\\ 
In the near and mid-infrared domain, the major contribution of the flux of hot Jupiter-like exoplanets is the absorbed, then re-emitted, component at the thermal equilibrium. In purely quantitative terms, the equilibrium temperature of such a planet approximately ranges from 500 to 1500 K. For the closest objects, the corresponding flux ratio between the planet and the star is about $10^{-5}$ to $10^{-4}$ in J band, $10^{-4}$ to $10^{-3}$ in K band, and several $10^{-3}$ in N band. In addition these planets exhibit several characteristics that we will detail in Section 4.1. \\
In this context let us consider a stellar system with a planet. We respectively note I$_*(\lambda)$ and I$_{\rm pl}(\lambda)$ the monochromatic flux of the two components separated by an angular distance $\boldsymbol{\alpha_{\rm pl}}$. The angular position of the center of the star is $\boldsymbol{\alpha_{*}}$. We first assume that the spatial location and extension of each component is independent of the wavelength. The star appears as a disc of constant angular diameter ${\rm D}_*$, defined by the normalized function $\Pi(\frac{\boldsymbol{\alpha}}{{\rm D}_*})$ equal to one inside the disc and zero outside. The planet appears as a point-like source defined by $\delta(\boldsymbol{\alpha})$. Both spatial distributions are respectively weighted by I$_*(\lambda)$ and I$_{\rm pl}(\lambda)$. The chromatic brightness distribution of this system is :
\begin{mathletters}
\begin{equation}
{\rm O}_{\rm syst}(\boldsymbol{\alpha},\lambda)=I_*(\lambda)\:\Pi(\frac{\boldsymbol{\alpha}-\boldsymbol{\alpha_*}}{{\rm D}_*})+I_{\rm pl}(\lambda)\:\delta(\boldsymbol{\alpha}-(\boldsymbol{\alpha_*}+\boldsymbol{\alpha_{\rm pl}})).
\end{equation}
The Fourier transform of the brightness distribution of the system in the plane of spatial frequency $\boldsymbol{\omega}$(u,v), covered by a single-baseline interferometer, is :  
\begin{equation}
\hat{O}_{\rm syst}({\rm \boldsymbol{u}},\lambda)=\kappa_*({\rm u})I_*(\lambda)e^{i2\pi \boldsymbol{{\rm u}\cdot \alpha_*}}+I_{\rm pl}(\lambda)e^{i2\pi \boldsymbol{{\rm u}\cdot (\alpha_*+\alpha_{\rm pl})}}.
\end{equation}
The u-axis is chosen along the interferometric baseline defined above, and takes the value $\boldsymbol{{\rm u}}=\frac{\boldsymbol{{\rm B}}}{\lambda}=\boldsymbol{\gamma_1}-\boldsymbol{\gamma_2}$. The intrinsic visibility of the star is $\kappa_{*}({\rm u})=C_{*}({\rm u})\:e^{i\Phi_{\rm star}}=2\frac{J_1(\pi{\rm D_*}{\rm u})}{\pi{\rm D_*}{\rm u}}$, with $J_1$ the Bessel function of the first kind of order one and ${\rm D_*}$ the stellar angular diameter. Since we are in a regime where the star is partly resolved or even unresolved by the interferometer, the stellar complex visibility is always real and positive, with $\kappa_*({\rm u})=C_{*}({\rm u})$ (or identically $C_*(\lambda)$) and $\Phi_{\rm star}=0$. Otherwise, if the star was resolved, $\Phi_{\rm star}$ could be equal to $\pi$.\\
In principle, visibility and phase carry complementary informations, and both might be useful for detecting and characterizing extrasolar planets. Nevertheless, considering the current possibilities of calibration of both quantities in terms of instrumental and atmospheric effects \citep{2006MNRAS.367..825V}, we prefer focussing our attention on the differential phase observable. 
The fringe phase is the argument of $\hat{O}_{\rm syst}({\rm \boldsymbol{u}},\lambda)$ \footnote{The $\lambda$ dependency of $\Phi_{\rm pl}(\lambda)$, $\Phi_{\rm *}(\lambda)$ $I_{\rm pl}(\lambda)$, ${\rm I}_{\rm *}(\lambda)$ and $C_*(\lambda)$ has been removed in Eq.(8c), Eq.(8d), and Eq.(8e), for lightening the notations.}:
\begin{equation}
\Phi_{\rm syst}(\lambda)=\arctan\frac{C_*I_*\sin(\Phi_*)+I_{\rm pl}\sin(\Phi_*+\Phi_{\rm pl})}{C_*I_*\cos(\Phi_*)+I_{\rm pl}\cos(\Phi_*+\Phi_{\rm pl})},
\end{equation}
with $\Phi_{\rm pl}(\lambda)=2\pi\boldsymbol{u}\cdot\boldsymbol{\alpha_{\rm pl}}$ and $\Phi_*(\lambda)=2\pi\boldsymbol{u}\cdot\boldsymbol{\alpha_*}$. Identically to the unresolved case, we chose to define $\Phi_{\rm syst}(\lambda)$ as a zero-mean differential phase in order to remove the ambiguity related to the measurement of the phase fringes.  
We also generalize the expression of $\Phi_{*}(\lambda)$ by taking into account other sources like instrumental instabilities or atmospheric perturbations that may contribute to this phase term. We thus have : $\Phi_*(\lambda)=\frac{2\pi}{\lambda}\delta_*$, where $\delta_*$ is the same residual piston shown in Fig. 1 and described in the unresolved case.\\
The coherent flux is the modulus of $\hat{O}_{\rm syst}({\rm \boldsymbol{u}},\lambda)$:
\begin{equation}
\rho_{\rm syst}(\lambda)=\sqrt{C_*^2I_*^2+I_{\rm pl}^2+2C_*I_*I_{\rm pl}\cos(\Phi_{\rm pl})}.
\end{equation}
 Finally, replacing $\Phi_{12}$ and $\rho_{12}$ by $\Phi_{\rm syst}(\lambda)$ and $\rho_{\rm syst}(\lambda)$ in Eq.(5e), allows us to deduce the parasitized phase of a stellar system with a hot Jupiter, $\chi_{\rm syst}(\lambda)$ :
{\small
\begin{equation}
\chi_{\rm syst}(\lambda)=\arctan\frac{t_{12}(1-\epsilon''^2)[C_*I_*\sin(\Phi_*)+I_{\rm pl}\sin(\Phi_*+\Phi_{\rm pl})]}{\epsilon''(1+t_{12}^2)(I_*+I_{\rm pl})+t_{12}(1+\epsilon''^2)[C_*I_*\cos(\Phi_*)+I_{\rm pl}\cos(\Phi_*+\Phi_{\rm pl})]}.
\end{equation}}The parasitized phase, $\chi_{\rm syst}(\lambda)$, contains the intrinsic and parasitic contributions of the phase. We thus consider $\chi_{\rm syst}(\lambda)-\Phi_{\rm syst}(\lambda)$ as the parasitic phase since it represents the parasitic phase contribution added to the intrinsic phase of the source, $\Phi_{\rm syst}(\lambda)$.  
This parasitic term clearly depends on $\Phi_*(\lambda)$ and $\Phi_{\rm pl}(\lambda)$, which appears both in $\Phi_{\rm syst}(\lambda)$ and $\chi_{\rm syst}(\lambda)$ and does not disappear during the subtraction. The parasitic phase contribution appears to be object-dependent. 
\end{mathletters}     
\section{Quantitative study of the 'unresolved source' case}
 In this Section, we evaluate the impact of parasitic fringes for a tilted wavefront originating from an unresolved source having a projected angular distance to the zenith different from zero. This very common situation can imply an uncorrected residual path difference between both telescopes. The principle of astrometric measurement of the position of an off-centre star via the phase of fringes may be affected by this situation.\\ 
The L band is chosen in this study since it corresponds to the most favourable band for detection of several astrophysical sources including hot Jupiter-like exoplanets \citep{vannier2003}. The parameters involved here are the parasitic flux factor $\epsilon"^2$, which describes the percentage of parasitic flux, and the piston $\delta$. This piston will be hereafter expressed in fraction of the L band central wavelength, namely 3.5 $\mu$m.\\
A typical baseline of 100 meters, and a range of piston values, from $\lambda/500$ to $\lambda/5$, are considered for calculating the phase signal of the source $\Phi_{\rm point}(\lambda)$. In this range, two piston values are especially of interest : $\lambda/500$ and $\lambda/30$; they approximately correspond to the typical specifications of piston correction achieved by a fringe tracking device in the case of space and groud-based interferometers (respectively 2 and 100 nm). In parallel, for each piston value, a range of parasitic flux factors, from 0 to 10 \%, is used for evaluating the parasitized phase signal $\chi_{\rm point}(\lambda)$.\\  
Finally we calculate ${a}_{\:\Phi}={\rm max}(\Phi_{\rm point}(\lambda))-{\rm min}(\Phi_{\rm point}(\lambda)$), and ${a}_{\:\rm par*}={\rm max}(\chi_{\rm point}(\lambda)-\Phi_{\rm point}(\lambda))-{\rm min}(\chi_{\rm point}(\lambda)-\Phi_{\rm point}(\lambda))$. These quantities represent the amplitude, on the extent of L band, of the intrinsic and parasitic phases. They are used for comparing the overall amplitude of the intrinsic and parasitic phases in L band, instead of comparing both phases at all the wavelengths. As the value of the parasitic phase depends on two independent parameters and varies with respect to the wavelength, this overall amplitude gives a typical value of the expected measurable phase signal in L band. Then this typical value can be plotted in a readable 2D representation as a function of the parasitic flux level. In Fig. 3, we plotted these two amplitudes in function of $\epsilon"^2$, considering different values of the piston parameter. We can notice that for all the piston values, the amplitude of the parasitic phase, $a_{\:\rm par*}$, is a smooth crescent function of the parasitic flux factor. 
Regarding the evolution of the intrinsic and parasitic phase amplitudes as a function of the piston value, it appears that $a_{\:\rm par*}$ always lies below 10$^{-2}$ radians while the amplitude of the phase signal of the star, $a_{\Phi}$, increases up to 0.5 radians.\\
Several conclusions can be drawn from this, depending on the context and the objectives of the observation :
\begin{itemize}
\item First we place in a classical interferometric context where the piston error has to be corrected before the addition of frames. A reasonable parasitic flux contribution of about 1\%, and an amount of uncorrected piston within the specifications of fringe tracking ($\lambda/500$ and $\lambda/30$ in L band) are assumed. In this case, the additional parasitic phase amplitude always lies below the intrinsic signal, and ranges from one third to one fifth of this intrinsic phase amplitude.
\item Second we consider an 'astrometric' approach, where the aim is to measure the phase signal created by an off-centre star, with respect to an on-axis one. It appears here that the measurement of the intrinsic phase amplitude is not degraded by the additional parasitic phase contribution especially for large piston values. The ratio between the intrinsic and parasitic phase amplitudes ranges from 3 to 13.
\end{itemize}     

 \section{Hot Jupiter signal and magnitude of parasitic fringes in L band}
  \subsection{Characterization of planetary fluxes}
  The fine study of the atmosphere of hot Jupiter-like exoplanets requires a characterization of their spectra since they may show significant differences from black body models. 
Contrary to the giant planets of our solar system, hot Jupiters are subject to an extremely intense stellar irradiation. Consequently, the temperature-pressure profile of the atmosphere is modified and a radiative area is going to develop and govern the cooling and contraction of the planet interior \citep{1996ApJ...459L..35G}. These structural changes of the atmosphere, compared to an isolated planet, strongly depend on how the stellar irradiation penetrates the planet interior. The albedo and the sources of opacity, related to the presence of different types of dust, are two important parameters allowing us to characterize and classify the theoretical spectra of hot Jupiter-like exoplanets.\\
\citet{2000ApJ...538..885S} distinguished five categories of theoretical spectra classified according to a range of effective temperatures, from T$\leq$150K to T$\leq$1500K. The albedo of the objects belonging to these classes, exhibits similar features directly related to the chemical composition of the planet.
In our case, the main criterion we use is the presence or absence of dust in the high layers of the atmosphere. For example, in a 'condensed' atmosphere model, dust has been settled down by the gravitational field or by the precipitations linked to the condensation and the formation of clouds.  
From a global point of view, this hypothesis is presumably more realistic for substellar objects with low temperatures, including the irradiated giant planets \citep{2000ARA&A..38..337C}.
Therefore, for modelling and estimating the variability of the parasitic phase produced during the observation of a hot Jupiter-like extrasolar planet, we use eight "condensed" synthetic spectra describing the planetary flux $f_{pl}(\lambda)$. They have been extracted from \citet{2001ApJ...556..885B}. Each spectrum corresponds to a given angular separation and intrinsic temperature T$_{\rm int}$ of the planet, which is irradiated either by a G2 or a M5 star respectively located 15 and 5 AU from the Sun. The flux of these stars is a simple black body law with temperatures of 5600 and 3000 K. 
Table 1 summarizes the values of the different parameters discriminating each synthetic spectrum.   
      
  \subsection{Parasitic phase estimation}
  In this Section, we estimate the impact of parasitic fringes in the context of the observation of different 'synthetic' stellar systems with a hot-Jupiter. 
  According to Sub-Section 3.3, the intrinsic phase $\Phi_{\rm syst}(\lambda)$, and the parasitized phase $\chi_{\rm syst}(\lambda)$,  
  depend on the flux ratio between the planet and the star, the separation $\boldsymbol{\rho}$ between both components, and the interferometric baseline $\boldsymbol{B}$; $\boldsymbol{B}$ being considered to be colinear with $\boldsymbol{\rho}$, and equal to 100m. $\Phi_{\rm syst}(\lambda)$ and $\chi_{\rm syst}(\lambda)$ are thus calculated for each spectrum. The ranges of values taken for the piston $\delta_*$ and the parasitic flux factor $\epsilon"^2$ are the same as those of the 'unresolved source' case.\\ 
 As detailed in Fig. 1, $\delta_*$ can be seen as the remaining piston uncorrected by the fringe tracking device. This residual piston creates a linear-like phase term, added to the intrinsic phase signature from the planet. Consequently, a further parasitic phase term will be added to the parasitic phase term due to the planet phase signal. It is therefore important to place this issue in the framework of a classical interferometric measurement since different contributions from the atmosphere and the instrument can affect the measurement of the interferometric phase and especially the first-order linear term via the residual piston (or piston error) $\delta_*$. Basically without any phase reference, the first-order linear term of the astrophysical signal is lost during the data reduction step estimating and removing this residual piston. This is the case with most current interferometers. Finally, the total phase, including the object phase and the parasitic contribution, contains only the higher order terms.  
Henceforth the effect of the parasitic interference has to be examined after removing the linear component of $\Phi_{\rm syst}(\lambda)$ and $\chi_{\rm syst}(\lambda)$, which consequently become 'unpistoned' phases.\\ 
For a clear illustration of this point, we plotted in Fig. 4 both 'unpistoned' quantities, $\Phi_{\rm syst}(\lambda)$ and $\chi_{\rm syst}(\lambda)$-$\Phi_{\rm syst}(\lambda)$, with respect to the wavelength, using the very smooth number 7 spectrum (see Table 1). $\chi_{\rm syst}(\lambda)$ has been calculated for two values of piston ($\lambda/500$ and $\lambda/30$) corresponding to the fringe tracking specifications, and a quite large parasitic flux factor of 10\%.   
First we can note a great similarity between both differential phases, highlighting the fact that the additional parasitic phase is an object-dependent quantity. It is quite low and anyway below the intrinsic term in both cases. However when the beams undergo a greater piston (here $\lambda/15$), the parasitic signal exceeds the intrinsic one and the phase signature from the planet is lost.\\ 
Identically to the 'unresolved source' case, we consider the amplitude in L band of the phase signature from the planet, ${a}_{\:{\rm syst}}={\rm max}(\Phi_{\rm syst}(\lambda))-{\rm min}(\Phi_{\rm syst}(\lambda))$, and of the corresponding parasitic phase, ${a}_{\:\rm par}={\rm max}(\chi_{\rm syst}(\lambda)-\Phi_{\rm syst}(\lambda))-{\rm min}(\chi_{\rm syst}(\lambda)-\Phi_{\rm syst}(\lambda))$. These quantities are calculated for a large range of $\delta_*$ values and are represented with respect to the parasitic flux factor $\epsilon''^2$ in Fig. 5. The $\lambda/500$ and $\lambda/30$ cases give a very similar parasitic phase amplitude that can only be distinguished for the spectra 5, 6 and 7. Here a detection is considered to be possible only if, for given values of $\delta_*$ and $\epsilon''^2$, the parasitic phase amplitude lies below the horizontal solid line representing the amplitude of the planet signal.  
Following this criteria, it appears that if typical fringe specifications are observed, the planet signal always lies above the parasitic phase amplitude; a hot Jupiter detection, with an intrinsic phase amplitude exceeding by a factor of three the parasitic one, is thus achieved with a parasitic intensity equal to 5\% of the total intensity.\\ 
However, when no fringe tracking is used, the expected constraint on the parasitic flux level appears to be much stronger. For instance, let us consider the case of a solar-like star (G2 type) and a 500K planet (spectra 5, 6). If the piston error is greater or equal to $\lambda/15$, a detection, with a planet phase amplitude standing at a factor three above the parasitic one, is achieved with a tolerance of not more than 0.01\% on the parasitic flux. On the contrary, the spectra 3 and 4 constitute very favourable cases. If the piston error lies below $\lambda/10$, a similar detection level admits a tolerance of 5\% on the parasitic flux. Therefore it appears that as soon as the fringe tracking specifications are relaxed or not precisely observed, the parasitic phase signal dramatically increases. In this case, planetary detection may be prevented if the parasitic flux level is not accurately monitored. Nevertheless in some cases where the flux ratio between the planet and the star is sufficiently large ($\approx$ $10^{-2}$ to $10^{-2}$), this assertion has to be moderated given that a non-respect of fringe tracking specifications is less influential. In those cases, the tolerance on the parasitic flux level is similar to the one achievable with a fringe tracker. \\
         
 \section{Discussion} 
 In this work we have described the phenomenon of parasitic interference occuring inside an interferometric device. This interference degrades the modulus and the phase of the complex visibility. Two further components are added to the intrinsic fringe pattern, the former assimilated to Young-like fringes and the latter assimilated to mirror-like fringes with respect to the intrinsic pattern. The phase and the amplitude of these 'parasitic' fringes depend on the piston between beams, including the piston due to the object position. In an 'object-image' approach, it would mean that the point spread function of the interferometer is no longer invariant by translation and that the object-image relation is anyway destroyed.\\
 In quantitative terms, we have shown that the parasitic phase is very sensitive to a piston between the interferometric beams. In fact a perfect unresolved source, observed at a projected angular distance from the zenith equal to zero, would not produce any parasitic phase. On the contrary, for an unresolved source undergoing an additional piston due to the atmosphere and the instrument, a parasitic flux factor of 1\% would create a parasitic phase reaching at most one third of the intrinsic phase amplitude.\\ 
 This effect is not negligible in the perspective of hot Jupiter-like planet observations. In this paper the feasibility study has been done in L band. We considered synthetic hot Jupiter spectra providing different flux ratios between the planet and the star, and different values of piston error within the specifications expected for fringe tracking devices. A hot Jupiter detection, with a planet phase signal three times larger than the parasitic phase amplitude, is possible if the parasitic flux reaches at most 5\% of the total incident flux.
Without any fringe tracking device, only hot-Jupiter like planetary systems with a quite large flux ratio between the planet and the star, for instance a dM5 star with a hot planet at 1000K, would admit such a tolerance of 5\%; however the values of piston error have to be equal or smaller than $\lambda/10$ in this case. For less advantageous planetary systems, the maximum admitted tolerance decreases up to 0.01\% for residual pistons equal to $\lambda/15$.\\
To summarize, we can see that the detection of astrophysical objects providing weak signatures in the interferometric phase, such as hot Jupiter-like extrasolar planets, requires careful attention to various fine instrumental effects such as parasitic fringes. This constitutes an important motivation for optimizing the design of future planet-detecting interferometers like the future ground-based instrument of the VLTI, MATISSE, or the future NASA spatial nulling interferometer, FKSI. In fact the calibration of this parasitic effect by reference star seems very difficult at this level of required precision. Moreover the use of a phase abacus is not possible since the parasitic phase is an object-dependent quantity. Nevertheless in the marginal case of an unresolved source, it appears that the parasitized coherent flux can be calibrated by the parasitized photometry. Therefore we propose some solutions to prevent parasitic interference effects, and which should be taken into account during the design stage of a future instrument.\\
We suggest the following design solutions :
\begin{itemize}
	\item To maintain the beams independence in certain parts of the instrument. For example, the left panel of Fig. 2 shows the crosstalk occuring if a spatial filtering plane is common to all beams. The solution would be to use an independent pinhole for each beam. 
	\item To separate the paths of each beam by a careful baffling everywhere it is possible inside the instrument.
	\item To transport the beams in a non-co-phased way except just before the recombination. These fixed path differences maintain the beams out of the coherence length in order to prevent any parasitic interference when cross-talk is occuring.
\end{itemize}
As a final conclusion, it appears that, up to now, little attention has been paid on the phenomenon of parasitic interference. This issue has been formalized here in a general multi-axial recombination scheme, with special attention on the differential phase.
This work has been conducted in the framework of the AMBER instrument \citep{2003EAS.....6..111P} and especially of its successor, the MATISSE instrument (\citet{2008SPIE.7013E..70L}; \citet{2008SPIE.7013E..94L}).\\
In the case of other instrumental configurations (co-axial scheme, nulling interferometry, ...), complementary work would need to be performed in order to define requirements adapted to a larger range of interferometers.

\acknowledgments
We thank our colleague, Pierre Antonelli, for the attention brought to 
the parasitic light issue, and for the fruitfull discussion motivated 
by the MATISSE instrument design study. We also thank the referee for his valuable comments that improved the article.
 
 
 \bibliography{articlefranges}

\begin{thebibliography}{15}
\providecommand{\natexlab}[1]{#1}
\providecommand{\url}[1]{\texttt{#1}}
\expandafter\ifx\csname urlstyle\endcsname\relax
  \providecommand{\doi}[1]{doi: #1}\else
  \providecommand{\doi}{doi: \begingroup \urlstyle{rm}\Url}\fi

\bibitem[{Barman} et~al.(2001){Barman}, {Hauschildt}, and
  {Allard}]{2001ApJ...556..885B}
T.~S. {Barman}, P.~H. {Hauschildt}, and F.~{Allard}.
\newblock {Irradiated Planets}.
\newblock \emph{ApJ}, 556:\penalty0 885--895, Aug. 2001.
\newblock \doi{10.1086/321610}.

\bibitem[{Bruning} et~al.(1974){Bruning}, {Herriott}, {Gallagher}, {Rosenfeld},
  {White}, and {Brangaccio}]{1974ApOpt..13.2693B}
J.~H. {Bruning}, D.~R. {Herriott}, J.~E. {Gallagher}, D.~P. {Rosenfeld}, A.~D.
  {White}, and D.~J. {Brangaccio}.
\newblock {Digital wavefront measuring interferometer for testing optical
  surfaces and lenses.}
\newblock \emph{Appl. Opt.}, 13:\penalty0 2693--2703, 1974.

\bibitem[{Chabrier} and {Baraffe}(2000)]{2000ARA&A..38..337C}
G.~{Chabrier} and I.~{Baraffe}.
\newblock {Theory of Low-Mass Stars and Substellar Objects}.
\newblock \emph{\araa}, 38:\penalty0 337--377, 2000.
\newblock \doi{10.1146/annurev.astro.38.1.337}.

\bibitem[{Elias} et~al.(2007){Elias}, {Harwit}, {Leisawitz}, and
  {Rinehart}]{2007ApJ...657.1178E}
N.~M. {Elias}, II, M.~{Harwit}, D.~{Leisawitz}, and S.~A. {Rinehart}.
\newblock {The Mathematics of Double-Fourier Interferometers}.
\newblock \emph{\apj}, 657:\penalty0 1178--1200, Mar. 2007.
\newblock \doi{10.1086/510878}.

\bibitem[{Fizeau}(1868)]{Fizeau1868}
H.~{Fizeau}.
\newblock \emph{C.R. Acad. Sc. Paris}, 66:\penalty0 932, 1868.

\bibitem[{Guillot} et~al.(1996){Guillot}, {Burrows}, {Hubbard}, {Lunine}, and
  {Saumon}]{1996ApJ...459L..35G}
T.~{Guillot}, A.~{Burrows}, W.~B. {Hubbard}, J.~I. {Lunine}, and D.~{Saumon}.
\newblock {Giant Planets at Small Orbital Distances}.
\newblock \emph{ApJL}, 459:\penalty0 L35+, Mar. 1996.
\newblock \doi{10.1086/309935}.

\bibitem[{Labeyrie}(1975)]{1975ApJ...196L..71L}
A.~{Labeyrie}.
\newblock {Interference fringes obtained on VEGA with two optical telescopes}.
\newblock \emph{ApJ}, 196:\penalty0 L71--L75, Mar. 1975.
\newblock \doi{10.1086/181747}.

\bibitem[{Lagarde, et al.}(2008)]{2008SPIE.7013E..94L}
S.~{Lagarde, et al.}
\newblock {MATISSE: concept analysis}.
\newblock In \emph{Society of Photo-Optical Instrumentation Engineers (SPIE)
  Conference Series}, volume 7013 of \emph{Society of Photo-Optical
  Instrumentation Engineers (SPIE) Conference Series}, July 2008.
\newblock \doi{10.1117/12.789391}.

\bibitem[{Lopez, et al.}(2008)]{2008SPIE.7013E..70L}
B.~{Lopez, et al.}
\newblock {MATISSE: perspective of imaging in the mid-infrared at the VLTI}.
\newblock In \emph{Society of Photo-Optical Instrumentation Engineers (SPIE)
  Conference Series}, volume 7013 of \emph{Society of Photo-Optical
  Instrumentation Engineers (SPIE) Conference Series}, July 2008.
\newblock \doi{10.1117/12.789412}.

\bibitem[{Michelson}(1920)]{1920ApJ....51..257M}
A.~A. {Michelson}.
\newblock {On the Application of Interference Methods to Astronomical
  Measurements}.
\newblock \emph{ApJ}, 51:\penalty0 257--+, June 1920.
\newblock \doi{10.1086/142550}.

\bibitem[{Petrov} and {Amber Consortium}(2003)]{2003EAS.....6..111P}
R.~G. {Petrov} and T.~{Amber Consortium}.
\newblock {The near infrared VLTI instrument AMBER}.
\newblock In G.~{Perrin} and F.~{Malbet}, editors, \emph{EAS Publications
  Series}, volume~6 of \emph{EAS Publications Series}, pages 111--+, 2003.

\bibitem[Schwider et~al.(1983)Schwider, Burow, Elssner, Grzanna, Spolaczyk, and
  Merkel]{Schwider:83}
J.~Schwider, R.~Burow, K.-E. Elssner, J.~Grzanna, R.~Spolaczyk, and K.~Merkel.
\newblock {Digital wave-front measuring interferometry: some systematic error
  sources}.
\newblock \emph{Appl. Opt.}, 22:\penalty0 3421--3432, 1983.
\newblock URL \url{http://ao.osa.org/abstract.cfm?URI=ao-22-21-3421}.

\bibitem[{Sudarsky} et~al.(2000){Sudarsky}, {Burrows}, and
  {Pinto}]{2000ApJ...538..885S}
D.~{Sudarsky}, A.~{Burrows}, and P.~{Pinto}.
\newblock {Albedo and Reflection Spectra of Extrasolar Giant Planets}.
\newblock \emph{ApJ}, 538:\penalty0 885--903, Aug. 2000.
\newblock \doi{10.1086/309160}.

\bibitem[Vannier(2003)]{vannier2003}
M.~Vannier.
\newblock \emph{Interférométrie et astrométrie différentielle chromatiques et
  observation de planètes extra-solaires géantes chaudes avec le VLTI et le
  NGST}.
\newblock PhD thesis, Laboratoire Universitaire d'Astrophysique de
  Nice/Université de Nice-Sophia Antipolis, mai 2003.

\bibitem[{Vannier} et~al.(2006){Vannier}, {Petrov}, {Lopez}, and
  {Millour}]{2006MNRAS.367..825V}
M.~{Vannier}, R.~G. {Petrov}, B.~{Lopez}, and F.~{Millour}.
\newblock {Colour-differential interferometry for the observation of extrasolar
  planets}.
\newblock \emph{MNRAS}, 367:\penalty0 825--837, Apr. 2006.
\newblock \doi{10.1111/j.1365-2966.2006.10015.x}.

\end{thebibliography}
 
 \begin{figure*}[!ht]
\centering
		\includegraphics[scale=0.6]{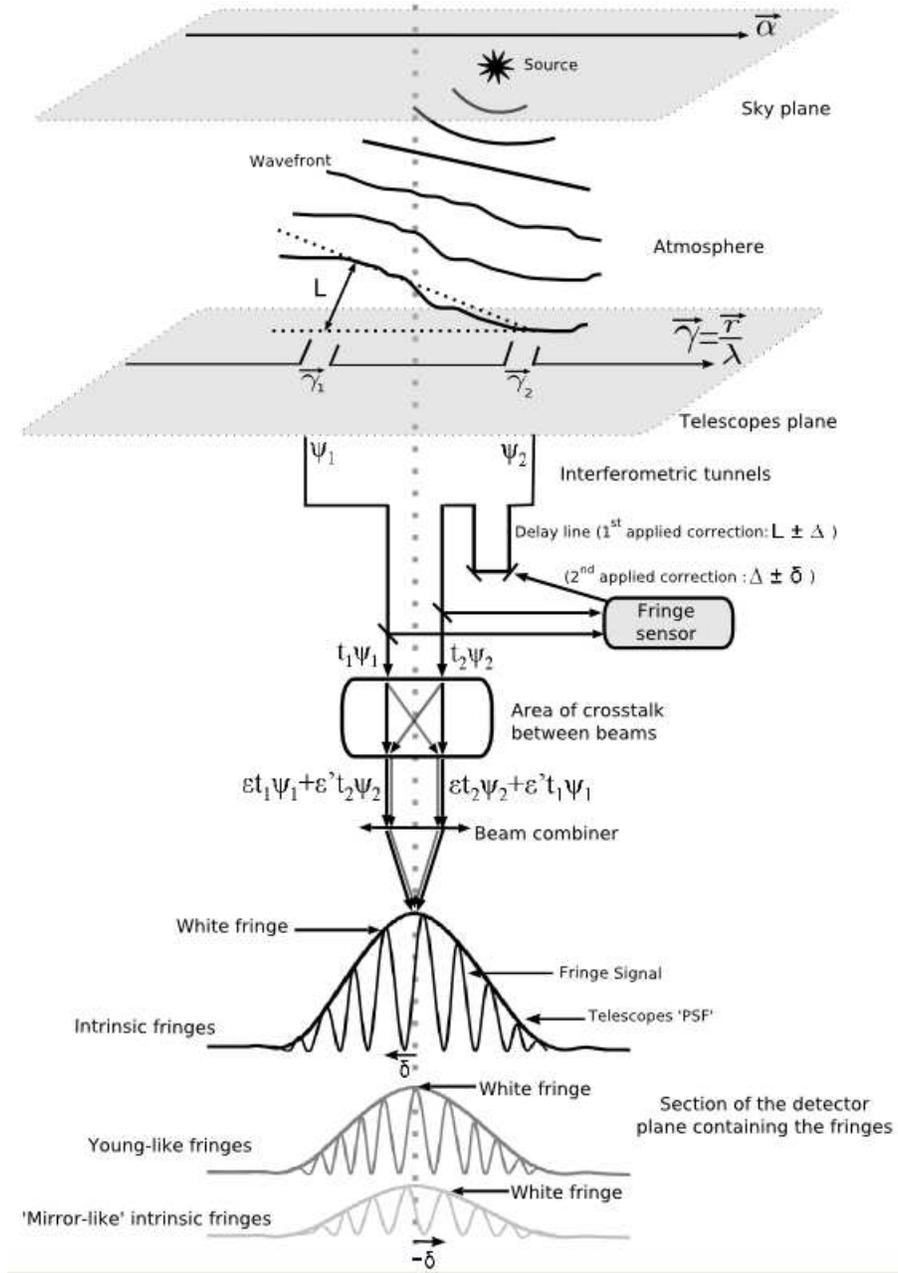}\hfill
		\caption{\footnotesize{Principle diagram representing the observation of an unresolved source by a 2-telescope interferometer. Fringes are obtained in the detector plane within a fixed fringe envelope corresponding to the point spread function of the telescopes. $\delta$ represents the residual achromatic piston that will affect the fringes within the signal envelope. The area where crosstalk between beams occurs, thus causing a parasitic interference, is simply represented by a 'module' located just before the beam recombiner. The fringe envelopes and patterns are normally overlapped at the same place on the detector but discriminated here for a matter of clarity.}}
\end{figure*}
 
 \begin{figure*}[!ht]
 \centering
		\includegraphics[width=77mm,height=54mm]{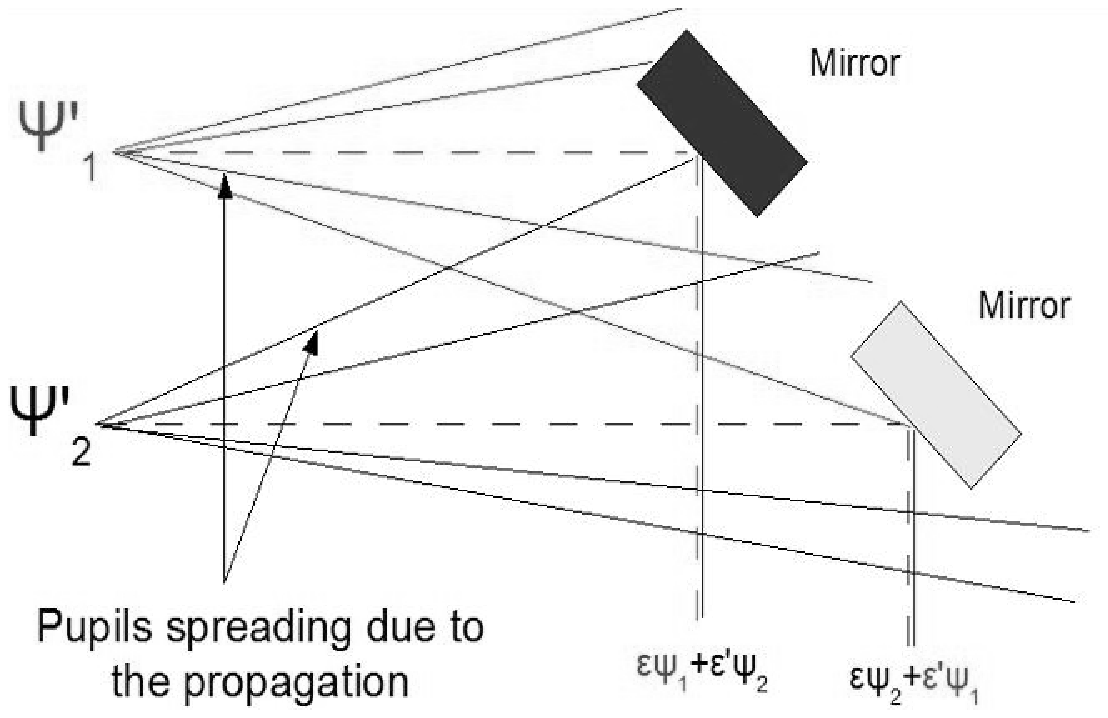}\hfill
		\includegraphics[width=77mm,height=54mm]{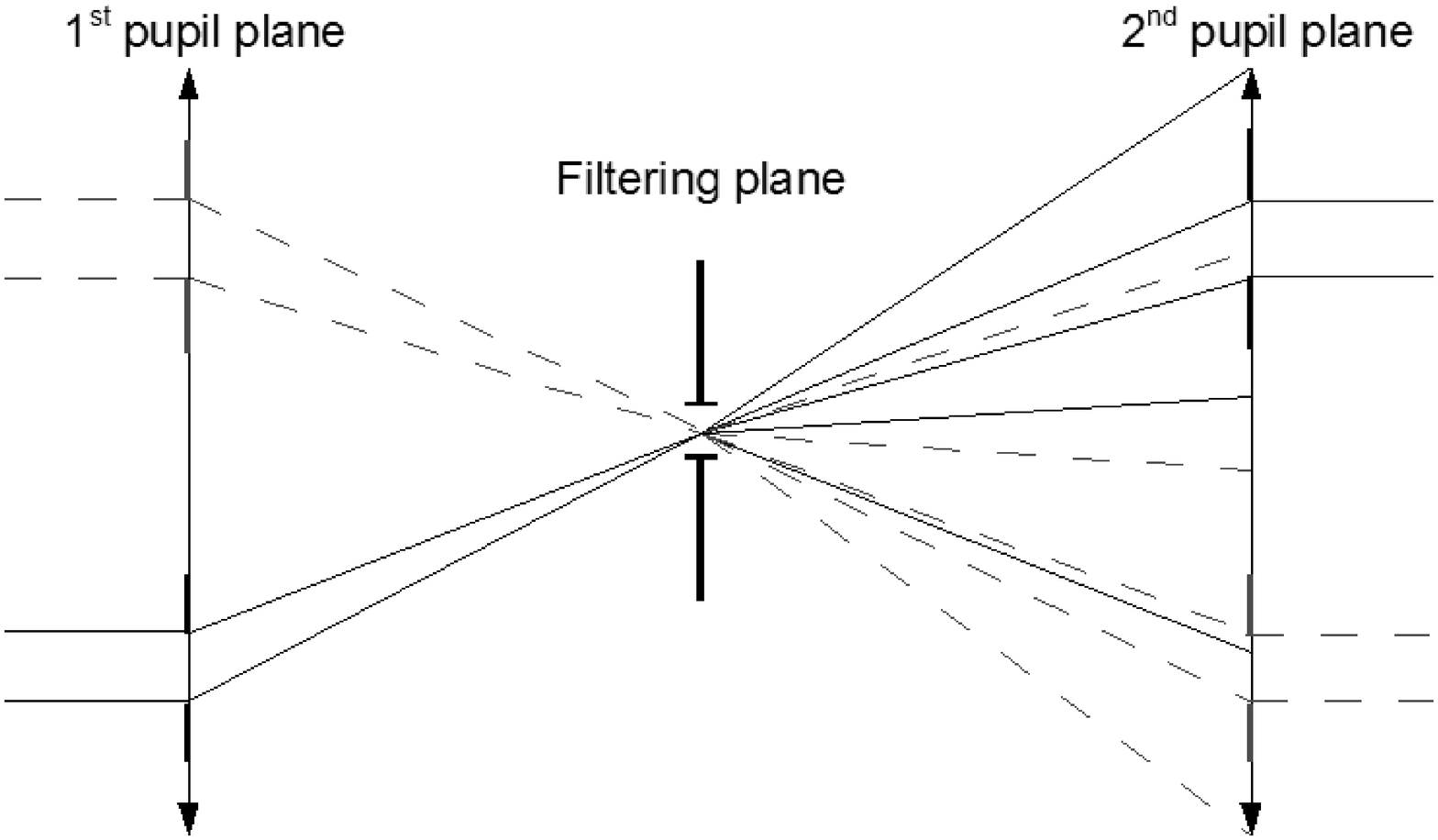}
		\caption{\small{Two examples of straylight and cross-talks between beams producing a parasitic interference : The left figure showing a beam overlapping due to diffusion effects, and the right one showing a cross-talk in a pupil plane due to the diffraction introduced by a common spatial filter.}}
\end{figure*}


\begin{figure*}[!t]
\centering
		\includegraphics[width=155mm,height=130mm]{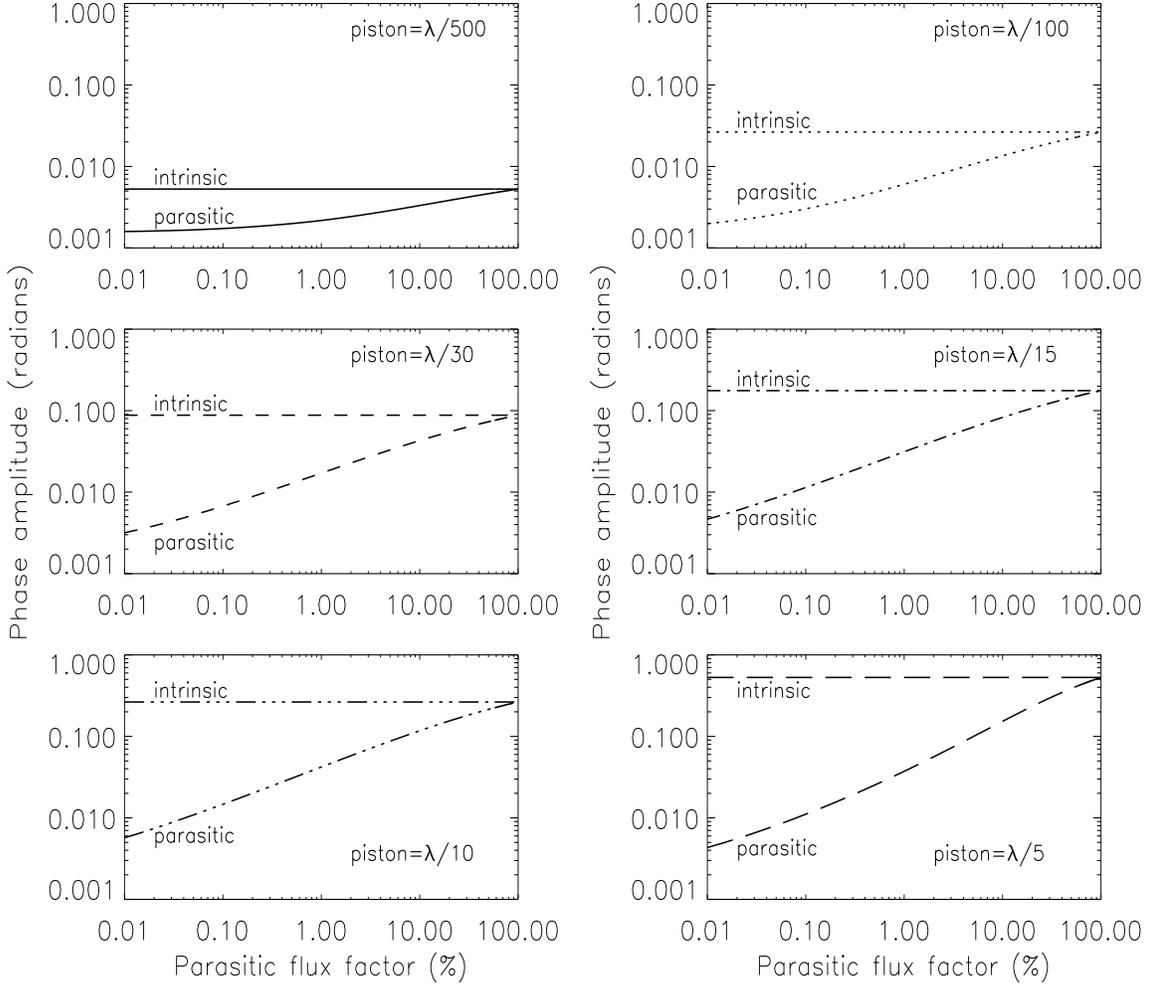}
		\caption{\footnotesize{Comparison between ${a}_{\:\Phi}$, the amplitude in L band of $\Phi_{\rm point}(\lambda)$ (noted 'intrinsic' in the figure), and ${a}_{\:{\rm par*}}$, the amplitude in L band of $\chi_{\rm point}(\lambda)-\Phi_{\rm point}(\lambda)$ (noted 'parasitic' in the figure), with respect to $\epsilon''^2$. Each panel corresponds to a piston value (noted $\delta$ in the paper) which is written in fraction of the central wavelength of the L band ($\lambda=3.5\mu m$). The $\lambda/500$ and $\lambda/30$ cases correspond approximately to the typical specifications that would be respectively achieved by a spatial and ground-based fringe tracking device ($\delta\approx5$nm and 100nm}).}
\end{figure*}

\begin{figure*}[!t]
\centering
		\includegraphics[width=85mm,height=70mm]{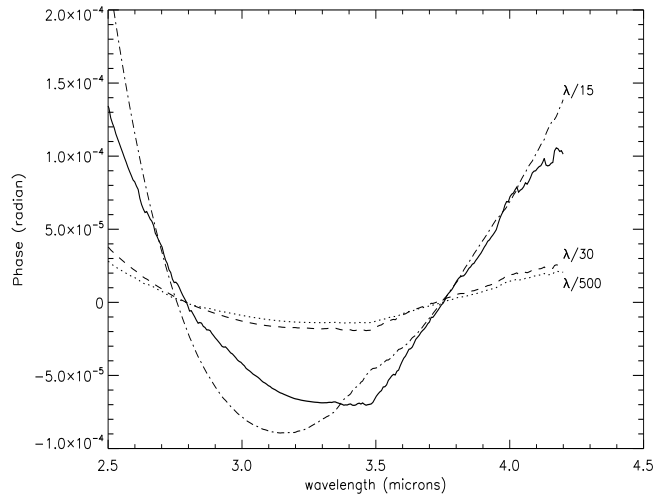}
		\caption{\footnotesize{Comparison between the 'unpistonned' intrinsic phase $\Phi_{\rm syst}(\lambda)$ (solid line) calculated from the number 7 spectrum, and different 'unpistonned' parasitic phases $\chi_{\rm syst}(\lambda)-\Phi_{\rm syst}(\lambda)$ calculated with different values of $\delta_*$ expressed in fraction of $\lambda=3.5\mu m$. The $\lambda/500$ and $\lambda/30$ cases correspond approximately to the typical specifications that would be respectively achieved by a spatial and ground-based fringe tracking device ($\delta_*\approx2$nm and 100nm). A parasitic flux factor of 10\% has been considered here.}}
\end{figure*}

\begin{figure*}
\vspace*{6mm}
\includegraphics[height=50mm,width=67mm,angle=0]{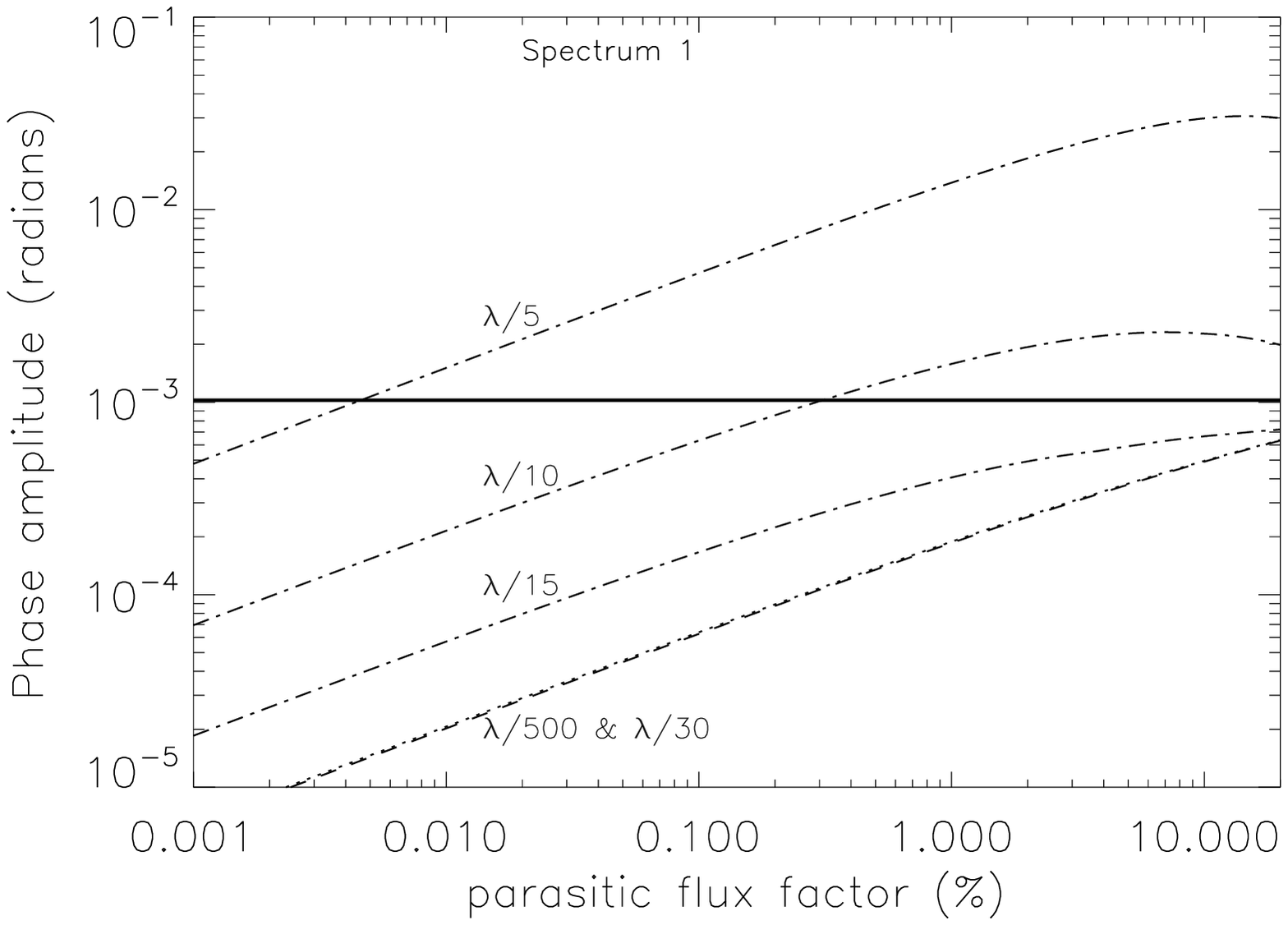}\hfill
\includegraphics[height=50mm,width=67mm,angle=0]{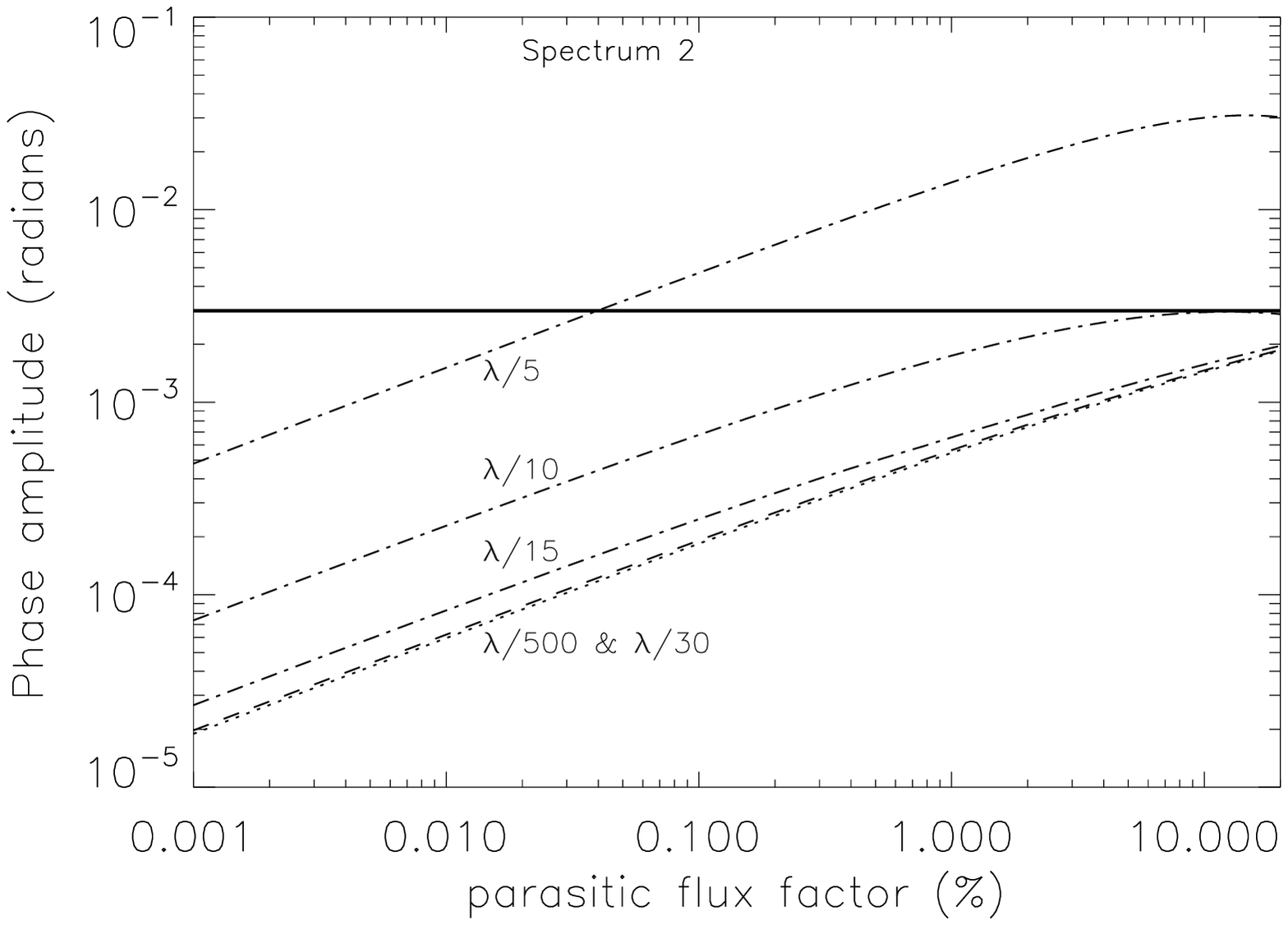}\hfill
\includegraphics[height=50mm,width=66mm,angle=0]{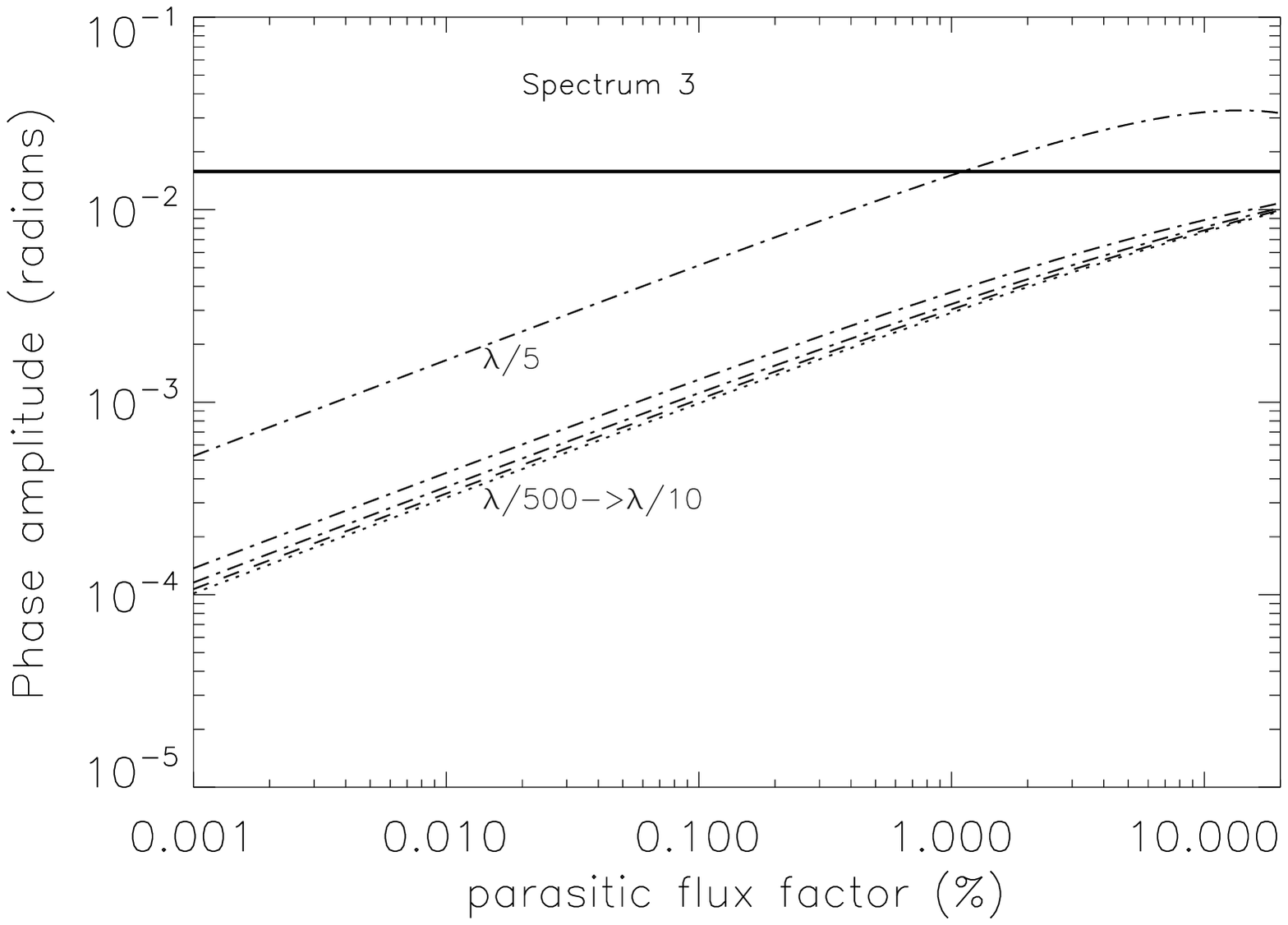}\hfill
\includegraphics[height=50mm,width=67mm,angle=0]{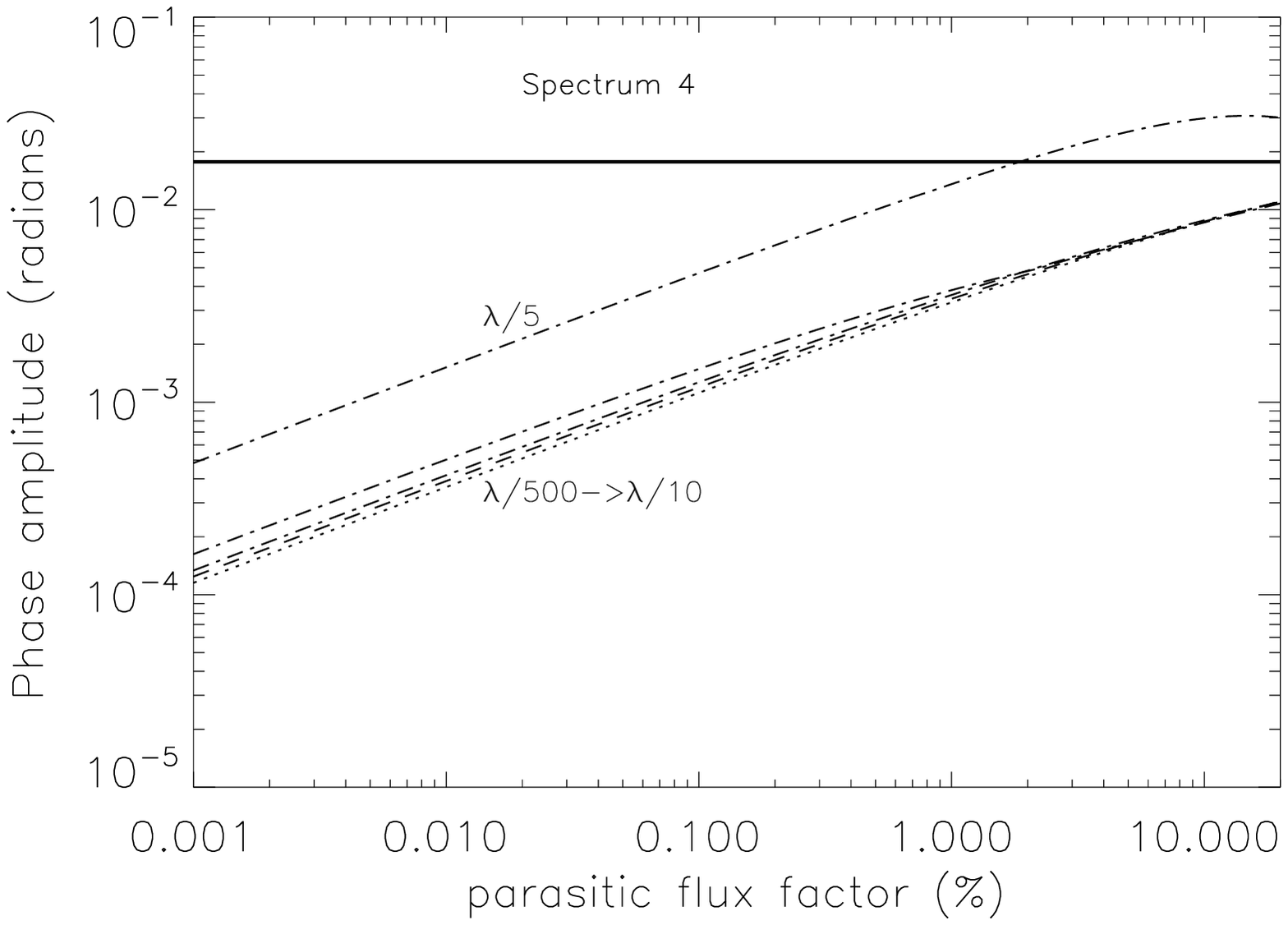}\hfill
\includegraphics[height=50mm,width=67mm,angle=0]{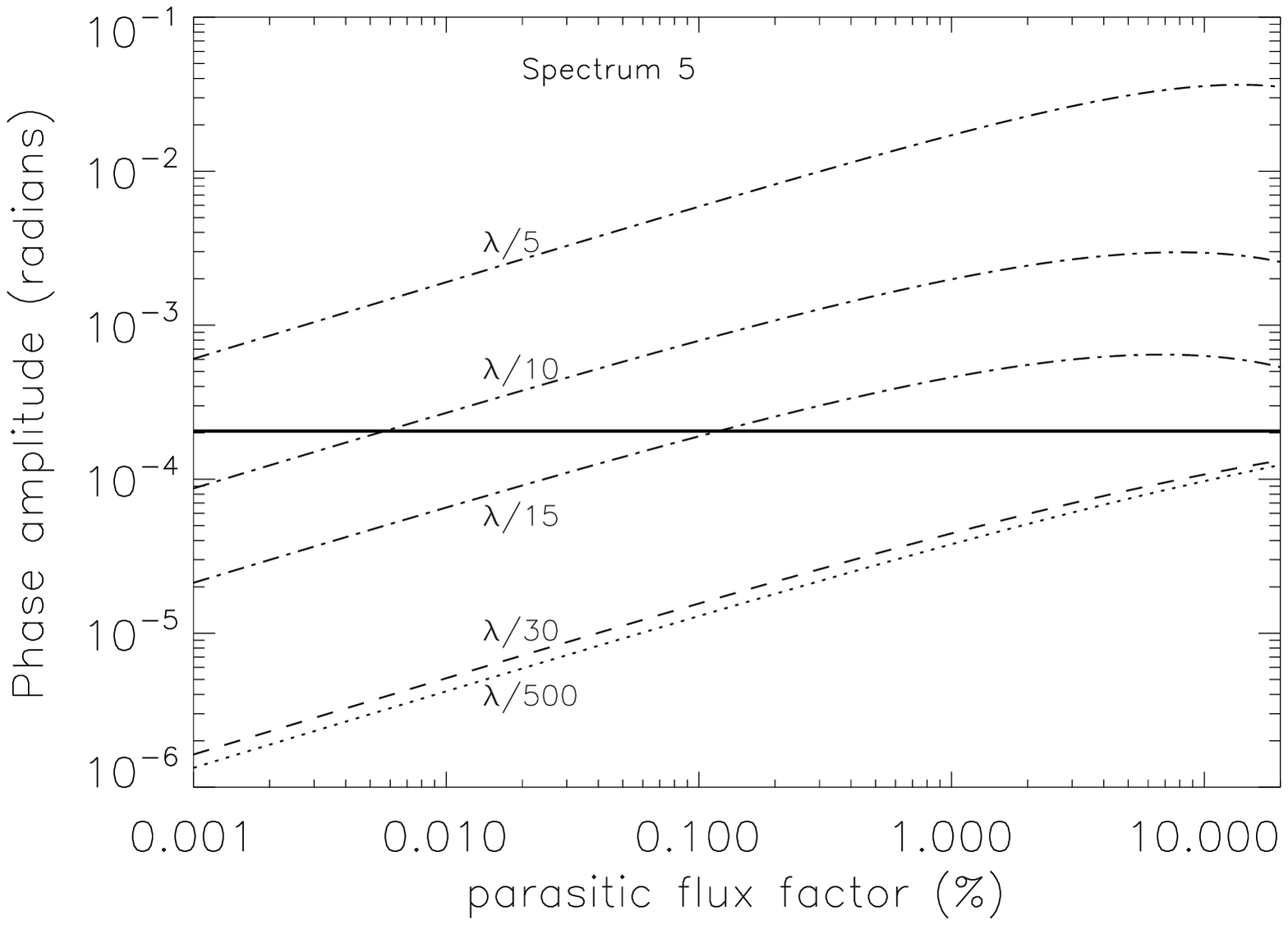}\ \ \ \ \ \ \ \ \ \ \ \ \ \ \ \ \ \ \ \ \ \ \ \ \hfill
\includegraphics[height=50mm,width=67mm,angle=0]{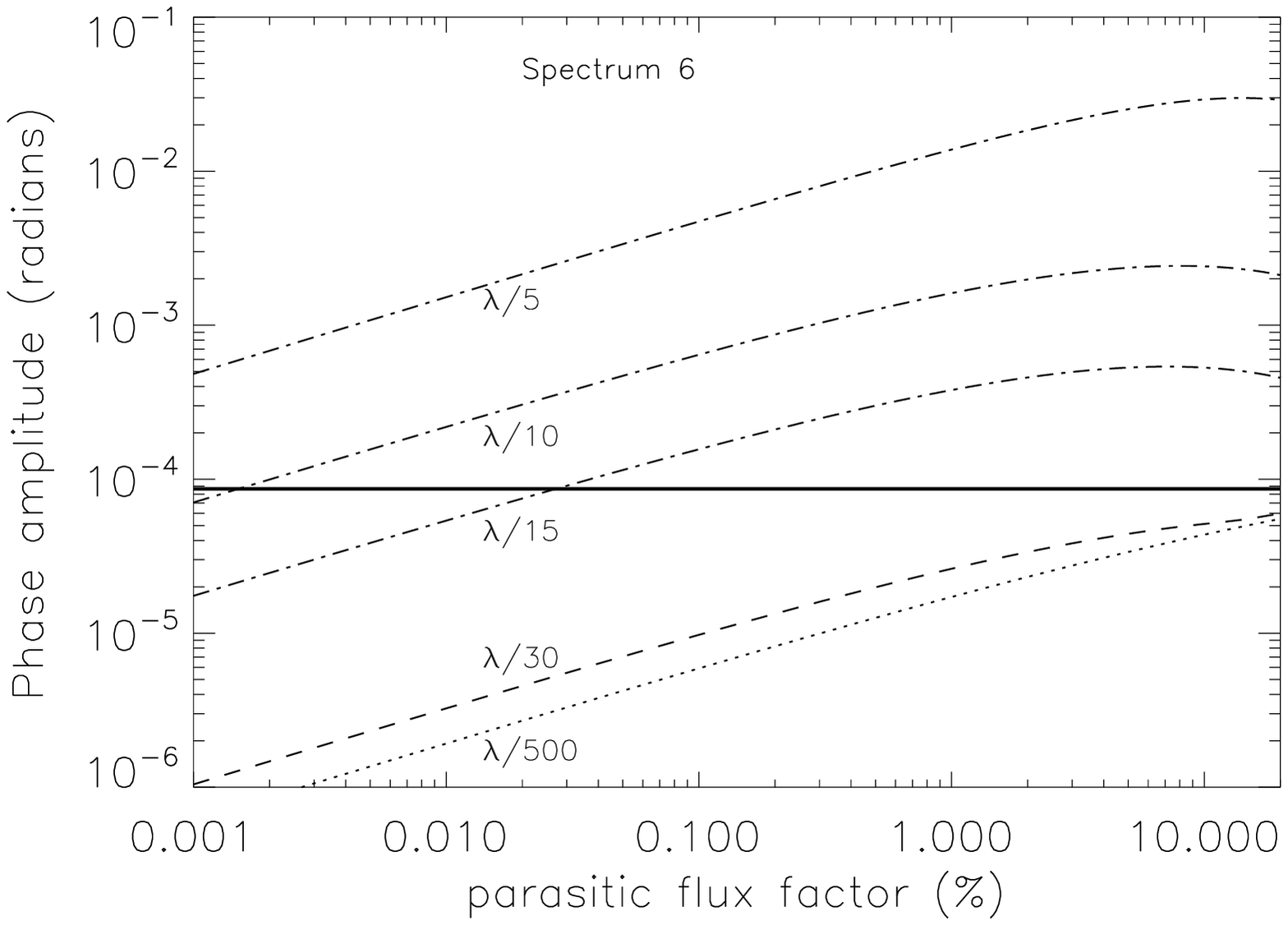}\hfill
\label{fig:multiple2}
\includegraphics[height=50mm,width=68mm,angle=0]{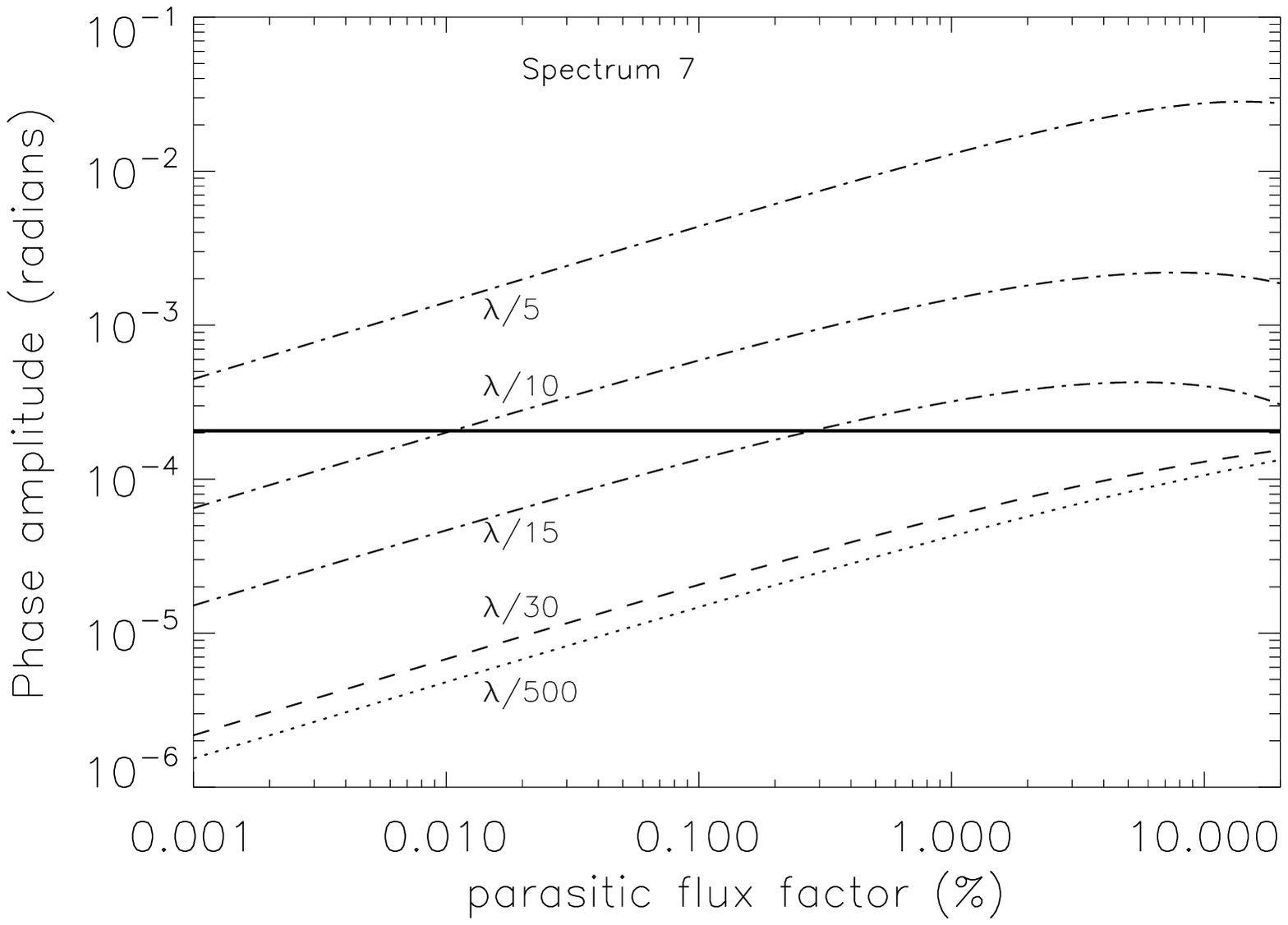}\hfill
\includegraphics[height=50mm,width=68mm,angle=0]{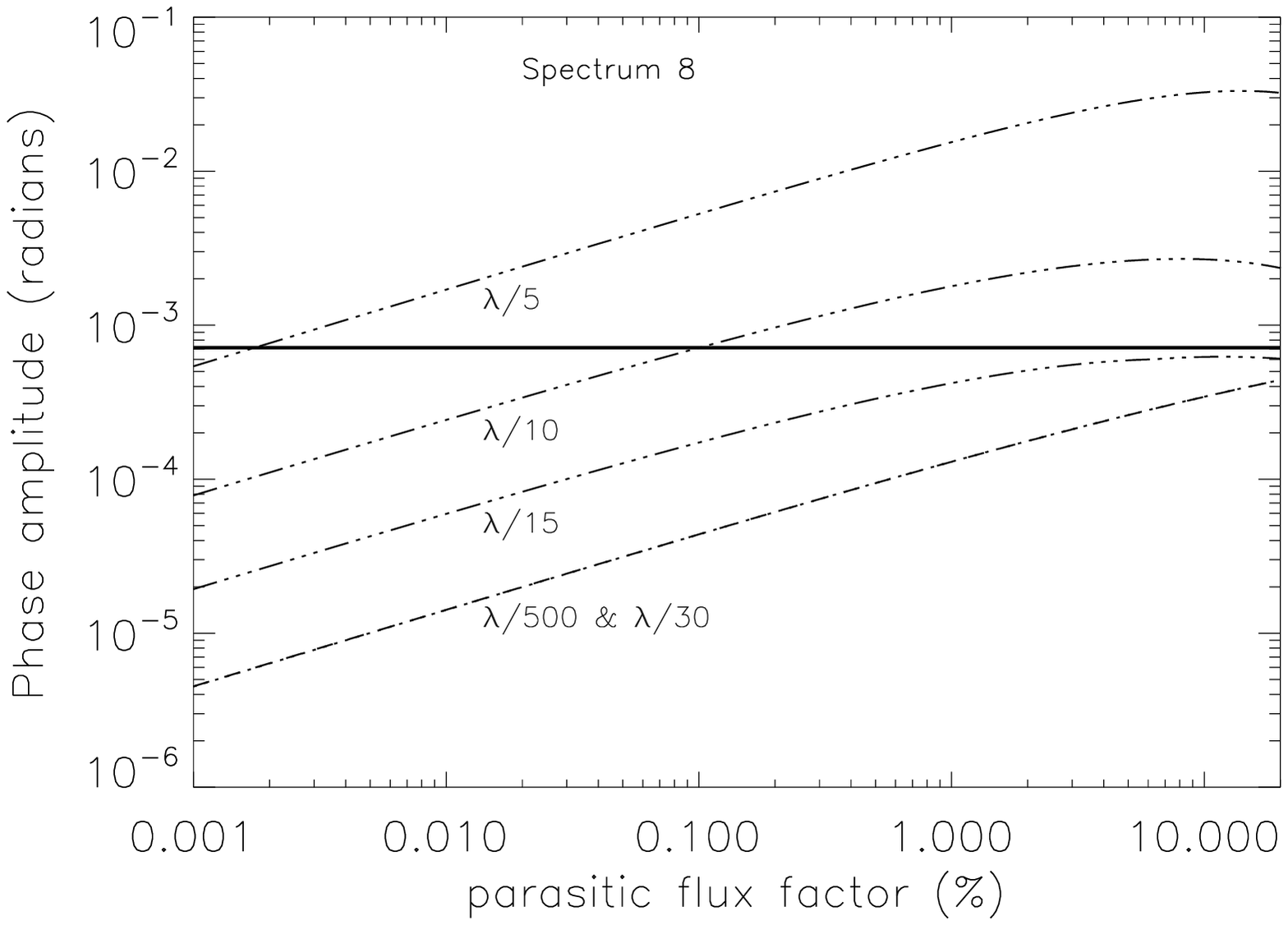}
\caption{\footnotesize{Amplitude, in L band, of the intrinsic phase of a stellar system with a hot Jupiter, $a_{\rm syst}$ (horizontal solid line), and amplitude in L band of the related parasitic phase, $a_{\rm par}$ (other dashed and dotted lines), plotted with respect to $\epsilon''^2$. Different values of $\delta_*$ ranging from $\lambda/500$ to $\lambda/5$ are considered, and each panel is related to a different hot Jupiter spectrum. The $\lambda/500$ and $\lambda/30$ cases correspond approximately to the typical specifications that would be respectively achieved by a spatial and a ground-based fringe tracking device ($\delta_*\approx2$nm and 100nm).}}
\end{figure*}

\begin{table}[!ht]
 \begin{center}
 \caption{Values of various parameters discriminating each phase spectrum extracted from \citet{2001ApJ...556..885B}.}
 \label{tab:table}
 \begin{tabular}[!h]{|c|c|c|c|}
 \hline
   & \textit{Type of star}& \textit {T$_{\rm int}$}& \textit{Separation star/planet} \\ 
 \hline
  spectrum 1& dM5& 500K& 0.05 AU\\
 \hline
  spectrum 2& dM5&  500K& 0.5 AU\\
 \hline
   spectrum 3& dM5&  1000K& 0.005 AU\\
 \hline
 spectrum 4& dM5 &  1000K& 0.1 AU\\
 \hline
  spectrum 5& G2 & 500K& 0.3 AU\\
  \hline
  spectrum 6 & G2 & 500K& 1 AU\\
 \hline
 spectrum 7 & G2 & 1000K& 0.05 AU\\
 \hline
 spectrum 8 & G2 &  1000K& 1 AU\\
 \hline
 \end{tabular}
 \label{tab:accreted}
 \end{center}
 \end{table}

\end{document}